\documentclass[bibyear]{aa}
\usepackage{graphicx}
\usepackage{amssymb}
\usepackage{tabulary} 
\usepackage{natbib,twoopt}

\begin{document}
   \title{An HST/COS legacy survey of intervening 
   Si\,{\sc iii} absorption in the extended gaseous halos 
   of low-redshift galaxies
   \thanks{Based on observations obtained with the NASA/ESA 
   Hubble Space Telescope, which is operated by the Space 
   Telescope Science Institute (STScI) for the Association of 
   Universities for Research in Astronomy, Inc., under NASA 
   contract NAS5D26555.}
}
%   \subtitle{}
   \author{ 
          P. Richter \inst{1,2},
          B.P. Wakker \inst{3},
	  C. Fechner \inst{1},
          P. Herenz \inst{1},
          T. Tepper-Garc{\'{\i}}a \inst{4},
	  \and
	  A.J. Fox \inst{5}
%         \fnmsep\thanks{Just to show the usage
%         of the elements in the author field}
          }
   \offprints{P. Richter\\
   \email{prichter@astro.physik.uni-potsdam.de}}

   \institute{Institut f\"ur Physik und Astronomie, Universit\"at Potsdam, 
             Karl-Liebknecht-Str.\,24/25, 14476 Golm, Germany
   \and 
   Leibniz-Institut f\"ur Astrophysik Potsdam (AIP), An der Sternwarte 16, 
   14482 Potsdam, Germany
   \and
   Department of Astronomy, University of Wisconsin-Madison,
   475 North Charter Street, \\ Madison, WI 53706, USA
   \and
   Sydney Institute for Astronomy, School of Physics, University of Sydney, 
   NSW 2006, Australia
   \and
   Space Telescope Science Institute, 3700 San Martin Drive, 
   Baltimore, MD 21218, USA
   }
   
   \date{Received 23/07/2015; accepted 10/03/2016}

%%%%%%%%%%%%%%%%%%%%%%%%%%%%%%%%%%%%%%%%%%%%%%%%%%%%%%%%
%%%%%%%%%%%%%%%%%%%%%%%%%%%%%%%%%%%%%%%%%%%%%%%%%%%%%%%%

\abstract
% context heading (optional)
% {} leave it empty if necessary
{}
% aims heading (mandatory)
{Doubly ionized silicon (Si\,{\sc iii}) is a powerful tracer of diffuse
ionized gas inside and outside of galaxies.
It can be observed in the local Universe in ultraviolet (UV) absorption
against bright extragalactic background sources.
We here present an extensive study of intervening Si\,{\sc iii}-selected absorbers 
and study the properties of the warm circumgalactic medium (CGM) around 
low-redshift ($z\leq0.1$) galaxies.}
% methods heading (mandatory)
{We analyzed the UV absorption spectra of 303 extragalactic 
background sources, as obtained with the Cosmic Origins Spectrograph (COS) on-board
the {\it Hubble Space Telescope (HST)}. 
We developed a geometrical model for the absorption-cross section of
the CGM around the local galaxy population and compared the observed
Si\,{\sc iii} absorption statistics with predictions provided by the model.
We also compared redshifts and positions of the absorbers with those of
$\sim 64,000$ galaxies using archival galaxy-survey data to
investigate the relation between intervening Si\,{\sc iii} absorbers
and the CGM.}
% results heading (mandatory)
{Along a total redshift path of $\Delta z\approx 24$, we identify $69$ intervening 
Si\,{\sc iii} systems that all show associated absorption from
other low and high ions (e.g., H\,{\sc i}, Si\,{\sc ii}, Si\,{\sc iv}, 
C\,{\sc ii}, C\,{\sc iv}).
We derive a bias-corrected number density of $d{\cal N}/dz($Si\,{\sc iii}$)=2.5\pm 0.4$ 
for absorbers with column densities log\,$N($Si\,{\sc iii}$)>12.2$, which 
is $\sim 3$ times the number density of strong Mg\,{\sc ii} systems at $z=0$.
This number density matches the expected cross section of a Si\,{\sc iii} absorbing
CGM around the local galaxy population with a mean covering fraction of 
$\langle f_c \rangle=0.69$.
For the majority ($\sim 60$ percent) of the absorbers, we identify possible 
host galaxies within $300$ km\,s$^{-1}$ of the absorbers 
and derive impact parameters $\rho<200$ kpc, demonstrating that the spatial 
distributions of Si\,{\sc iii} absorbers and galaxies are highly correlated.}
% conclusions heading (optional), leave it empty if necessary
{Our study indicates that the majority of Si\,{\sc iii}-selected
absorbers in our sample trace the CGM of nearby galaxies within their
virial radii at a typical covering fraction of $\sim 70$ percent.
We estimate that diffuse gas in the CGM 
around galaxies, as traced by Si\,{\sc iii}, contains substantially more 
(more than twice as much) baryonic mass than their neutral interstellar medium.
}

\titlerunning{Intervening Si\,{\sc iii} absorbers}

\maketitle

%%%%%%%%%%%%%%%%%%%%%%%%%%%%%%%%%%%%%%%%%%%%%%%%%%%%%%%%

\section{Introduction}

Galaxies at low and high redshift are
surrounded by massive gaseous halos. These halos are believed to 
be built up and fed by large-scale gas circulation processes, such as
accretion from the intergalactic medium (IGM), galactic winds, and outflows 
powered by star formation and active galactic nuclei, as well as 
minor and major mergers that result from the on-going hierarchical
formation and evolution of galaxies. Such gaseous material in extended
galaxy halos is nowadays often referred to as the circumgalactic medium
(CGM). We here define the CGM as diffuse neutral and ionized gas that 
is located within the virial radius of a galaxy, but outside of its (main) 
stellar body.

Observational and theoretical studies imply that the CGM
is a key component in the ongoing process of galaxy formation 
and evolution. Because the gas-consumption timescales of 
late-type galaxies are short compared to their lifetimes,
they must gain gaseous material from outside (e.g, from the IGM
and/or from merger processes) through the CGM to sustain their
ongoing star formation. The manner in which the gas makes its way 
from the IGM/CGM to the disks of late-type spirals is not well
understood, however. The classical picture of 
accretion (e.g., Rees \& Ostriker 1977; White \& Rees 1978; 
Birnboim \& Dekel 2003)  and the "hot-" and "cold-mode" scenarios
(Kere$\check{s}$ et al.\,2005) most likely are oversimplified, 
because the underlying physics that describe the 
large-scale flows of multiphase gas from the outer to the inner 
regions of a dynamically evolving galaxy is highly complicated
(e.g, Mo \& Miralda-Escude 1996; Maller \& Bullock 2004). Most likely, 
only state-of-the art hydrodynamical simulations that cover the 
necessary physics and that have the necessary spatial resolution 
to potentially provide realistic measures for the amount of gas
that is reaching the disk of late-type galaxies and for the timescales 
that it requires to do so (e.g., van\,de\,Voort et al.\,2011).

Observations that aim at studying the properties of the CGM are thus extremely
important to improve our understanding of gas-accretion processes and provide
constraints for numerical models.
What observations can provide is a characterization of the 
distribution of the different gas phases in the CGM, the total mass 
that it contains (under reasonable assumptions), 
and the connection between the properties of the CGM
and the properties of their host galaxies (e.g., morphology, luminosity,
star-formation rate, etc.). The gas densities in the CGM are low,
however, so that emission measurements of circumgalactic gas 
(e.g., in the X-ray regime) typically are limited to the very 
inner halo regions of nearby galaxies using current instruments 
(Anderson \& Bregman 2011; Dai et al.\,2012; Bogd{\'a}n et al.\,2013).
Fortunately, absorption-line measurements that use background active 
galactic nuclei (AGNs; throughout the following 
we use the abbreviation QSO to refer to the various classes of 
AGNs collectively) can access physical tracers at the relevant densities 
with extremely high sensitivity. For such absorption-line measurements 
the ultraviolet (UV) range is particularly important, as it contains
a large number of diagnostic transitions from low, intermediate and 
high ions of heavy elements and the Lyman series of neutral hydrogen.
Consequently, UV absorption-line observations with past and present 
space-based UV spectrographs such as the 
Far-Ultraviolet Spectroscopic Explorer ({\it FUSE}),
the Space Telescope Imaging Spectrograph (STIS) and
the Cosmic Origins Spectrograph (COS; the latter two instruments 
installed on the {\it Hubble Space Telescope}; \emph{HST})
provide a wealth of information on the physical
and chemical properties of the different gas phases in the
CGM of the Milky Way and other galaxies (e.g., Wakker et
al.\,1999, 2001; Sembach et al.\,2003; Richter et al.\,2001, 2009,
2011; Tripp et al. 2003; Fox et al. 2005, 2010; Collins et
al.\,2009; Wakker \& Savage 2009; Shull et al.\,2009; Prochaska et al.\,2011; 
Tumlinson et al.\,2011; Lehner et al.\,2013; Keeney et al.\,2013; Werk et al.\,2013;
Stocke et al.\,2014; Liang \& Chen 2014).

In two previous papers, we have studied the amount and distribution 
of neutral gas (the so-called high-velocity clouds) in the halos 
of low redshift galaxies via optical absorption spectroscopy and their 
contribution to the neutral-gas accretion rate in the local Universe 
(Richter et al.\,2011; Richter 2012). We here continue our long-term
strategy to study the circumgalactic medium in the local Universe
with an absorption-line survey that aims at characterizing the extent
and mass of diffuse ionized gas around low-redshift galaxies using
archival UV absorption-line data from \emph{HST}/COS.

This paper is organized as follows.
A short discussion on the importance of doubly ionized silicon as a
sensitive tracer for circumgalactic gas is presented in Sect.\,2.
In Sect.\,3 we describe the \emph{HST}/COS data acquisition, the
COS data reduction, the spectral analysis method, and the galaxy
data origin.
In Sect.\,4 we discuss the observed absorber properties, such as their
incident rate, their redshift distribution, the distribution of equivalent
widths and column densities.
In Sect.\,5 we model the expected absorption cross section of circumgalactic
gas using the local galaxy luminosity function.
Sect.\,6 deals with the observed absorber-galaxy connection in our
data sample.
The ionization conditions and the cosmological mass density of the
Si\,{\sc iii} absorbers are considered in Sect.\,7.
We discuss our results in Sect.\,8 and provide a summary of our 
study in Sect.\,9.

%%%%%%%%%%%%%%%%%%%%%% FIGURE 01 %%%%%%%%%%%%%%%%%%%%%%

\begin{figure}[t!]
\resizebox{1.0\hsize}{!}{\includegraphics{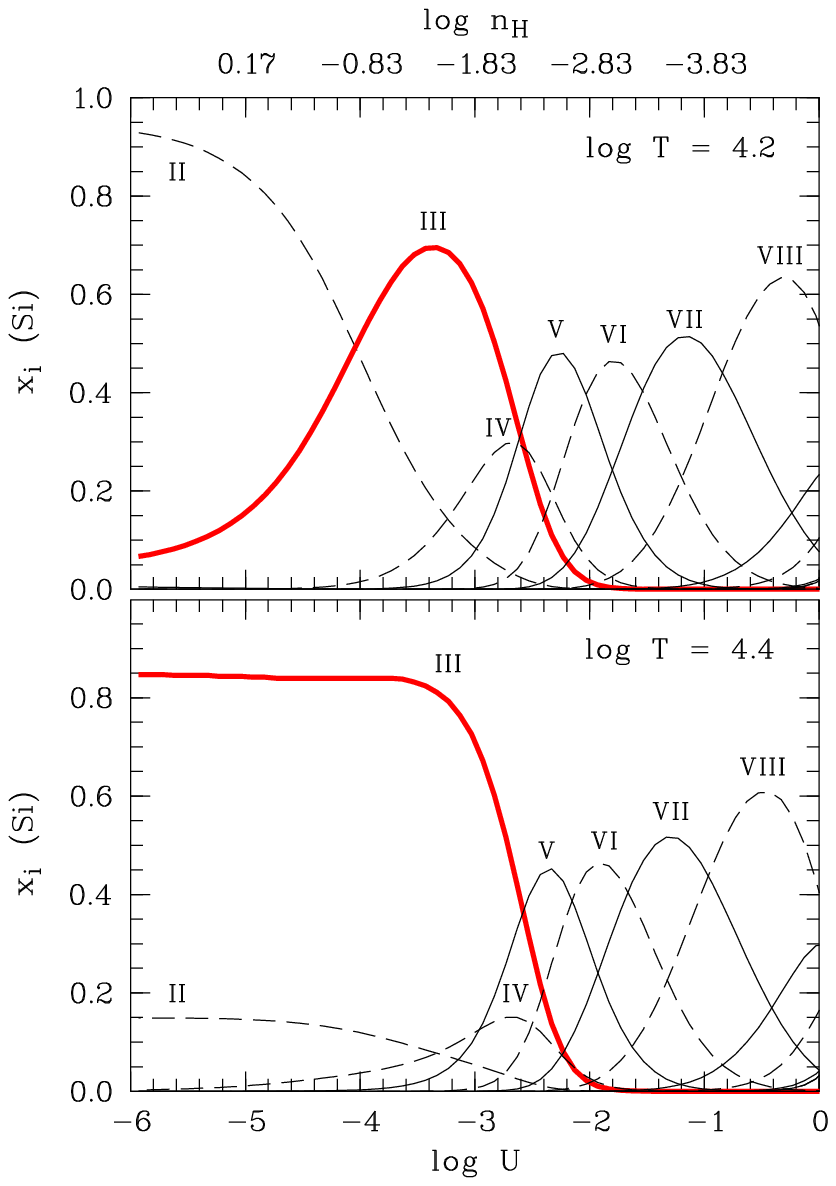}}
\caption[SiIII ionzation fractions]{
Ionization fractions of different Si ionization states for different gas densities
and temperatures, as calculated from a Cloudy ionization model (including
photoionization and collisional ionization) of
circumgalactic gas at $z=0$ using the local UV background.
The upper panel shows a model
for log $T=4.2$, the lower panel a model for log $T=4.4$.
The Si\,{\sc iii} ion represents the
dominant ionization state in the density range that is characteristic
for circumgalactic gas structures that are embedded in hot
coronal gas (log $n_{\rm H}=-2.0$ to $-3.5$).
}
\end{figure}

%%%%%%%%%%%%%%%%%%%%%%%%%%%%%%%%%%%%%%%%%%%%%%%%%%%%%%%%

\section{On the importance of Si\,{\sc iii} as a tracer of circumgalactic gas}

A particularly powerful transition in the UV to detect ionized interstellar, 
circumgalactic, and intergalactic gas in a wide range of 
physical conditions is that of doubly ionized silicon
(Si\,{\sc iii}) at $1206.500$ \AA\,(hereafter referred to 
as Si\,{\sc iii}\,$\lambda 1206$). This is because of the very
high oscillator strength of Si\,{\sc iii}\,$\lambda 1206$ of $f=1.63$ 
(Morton 2003) and the relatively large cosmic 
abundance of silicon (log (Si/H$)_{\sun}=-4.49$; Asplund et al.\,2009).
Doubly ionized silicon is a Mg-like ion with a closed-subshell
$3s^2$ valence-electron structure. The ionization-energy boundaries 
of the Si\,{\sc iii} ion are $16.35$ and $33.49$ eV (Morton 2003).
As a result, detectable Si\,{\sc iii} arises from both diffuse 
photoionized as well as collisionally ionized gas at moderate temperatures
($T<10^5$ K; see Shull et al.\,2009).

To demonstrate the diagnostic power of the Si\,{\sc iii}\,$\lambda 1206$
transition for the study of the CGM we show in Fig.\,1 the expected 
fractional abundance of the Si ions Si\,{\sc ii}$-$Si\,{\sc viii}
in gas at $z=0$ that is exposed to the local UV background. For the UV
background we use a modified version of the model by Haardt \& Madau (2012),
in which we re-scale the photoionization rate to a value of
log $\Gamma=-13.6$ to compensate for the apparent discrepancies between
the Haardt \& Madau model and recent observational results (see Kollmeier et al.\,2014;
Wakker et al.\,2015; Shull et al.\,2015).
The calculations (Fechner 2016, in prep.) 
are based on ionization models performed with the ionization code Cloudy 
(v13.03; Ferland et al.\,2013). 
As can be seen, doubly ionized Si represents the dominant 
observable ionization state of Si at densities and temperatures 
(log $n_{\rm H}=-2.0$ to $-3.5$, $T<10^5$ K)
that are typical for multiphase circumgalactic gas structures that are 
embedded in hot coronal gas (see Cen 2013).

In terms of quasar absorption-line systems, Si\,{\sc iii}\,$\lambda 1206$
absorption in the low-redshift Universe is expected to be detectable (if
not blended by other spectral features) in basically {\it all} damped 
Lyman $\alpha$ absorbers (DLAs; log $N$(H\,{\sc i}$)\geq 20.3$), 
sub-damped Lyman $\alpha$ absorbers (sub-DLAs; $19.0\leq$\,log $N$(H\,{\sc i}$)<20.3$),
Lyman-limit systems (LLS; $17.2\leq$\,log $N$(H\,{\sc i}$)<19.0$),
because these are metal-enriched, multiphase absorbers with high gas columns.
The majority of Si\,{\sc iii}-selected absorbers (throughout the 
following referred to simply as Si\,{\sc iii} absorbers), however,
are expected to arise in ionized metal systems at lower H\,{\sc i} 
column densities (log $N$(H\,{\sc i}$)<17.2$), where the minimum H\,{\sc i} column
density to detect Si\,{\sc iii} with {\it HST}/COS in solar-metallicity gas at 
moderate gas densities ($n_{\rm H}\sim 10^{-3}$ cm$^{-3}$) can be as low 
as log $N$(H\,{\sc i}$)=14$ (see Sect.\,7.1).

Low ions with lines in the observable UV, such as O\,{\sc i}, C\,{\sc ii}, 
and Si\,{\sc ii}, are only present in mostly neutral or slightly ionized gas 
(i.e., at high $N$(H\,{\sc i})), while high ions such as C\,{\sc iv} and O\,{\sc vi} 
predominantly trace highly-ionized gas at gas densities that 
are typically below $n_{\rm H}\sim 10^{-3}$ cm$^{-3}$.
A Si\,{\sc iii}-selected absorption-line survey, such as presented
in this paper, therefore is expected to be particularly sensitive to detect
metal-enriched gas in the inner and outer halos of
galaxies and to characterize its spatial distribution and
physical properties.

%%%%%%%%%%%%%%%%%%%%%% FIGURE 02 %%%%%%%%%%%%%%%%%%%%%%

\begin{figure*}[ht!]
\resizebox{1.0\hsize}{!}{\includegraphics{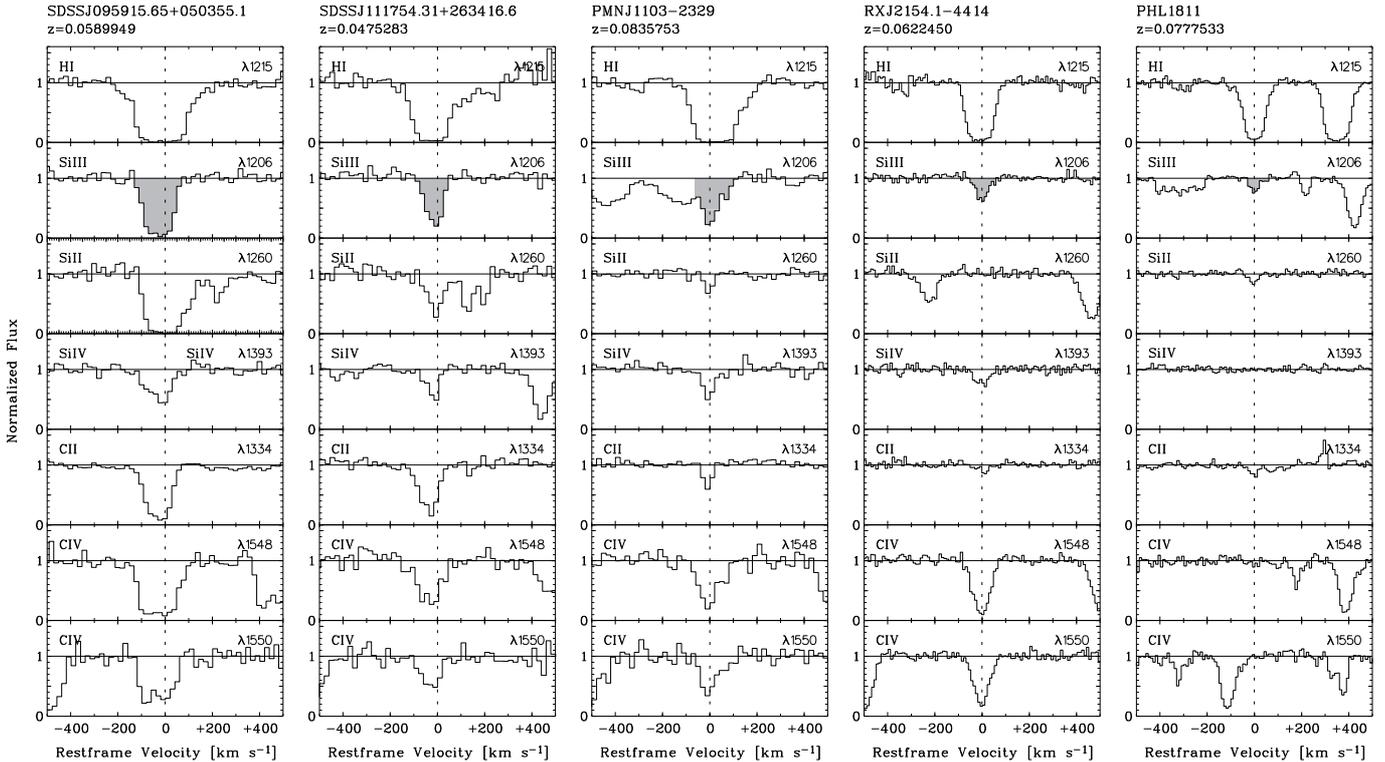}}
\caption[Velocity profiles of SiIII]{
Five examples for velocity profiles of Si\,{\sc iii}-selected absorption
line systems in our \emph{HST}/COS survey with decreasing
Si\,{\sc iii} $\lambda 1206$ absorption strengths (gray-shaded area;
from left to right).
Shown are the velocity profiles of H\,{\sc i}, Si\,{\sc iii}, Si\,{\sc ii},
Si\,{\sc iv}, C\,{\sc ii}, and C\,{\sc iv} (from top to bottom) 
plotted against the absorber's restframe velocity.
The names of the background QSOs and the absorption redshifts are indicated
above each panel. A complete set of velocity plots for all detected Si\,{\sc iii}
absorbers discussed in this paper is available in the Appendix.
}
\end{figure*}

%%%%%%%%%%%%%%%%%%%%%%%%%%%%%%%%%%%%%%%%%%%%%%%%%%%%%%%%

\section{Observations, data handling, and analysis method}

\subsection{COS spectra selection and data reduction}

For our study we make use of archival \emph{HST}/COS data that were
retrieved from the HST Science Archive at the Canadian 
Astronomy Data Centre (CADC). 
Because we aim at studying intervening Si\,{\sc iii} $\lambda 1206$ absorption 
in the IGM at $z\leq0.1$ together with corresponding H\,{\sc i} 
Ly\,$\alpha$ $\lambda 1215$ absorption we are primarily interested in 
the wavelength range between $1208$ and $1338$ \AA. 
This range is covered by the COS G130M grating which operates between 
$\lambda =1150-1450$ \AA, providing a spectral resolution of 
$R\approx 15,000-20,000$ 
(corresponding to an instrumental FWHM of $15-20$ km\,s$^{-1}$, while
the native pixel size is $2$ km\,s$^{-1}$; Green et al.\,2012; Debes et al.\,2016).
Using the CADC web interface we searched for all publicly available COS data from 
all types of extragalactic point sources such as the various types of AGN and 
galaxies that were observed with the COS G130M grating. 
By the end of February 2014 we found (and downloaded) G130M data sets 
for 552 extragalactic sightlines. 
Whenever G160M data (covering the range $\lambda=1405-1775$ \AA\, including
the important C\,{\sc iv} doublet at $1548.2,1550.8$ \AA) were available for 
these sightlines, we retrieved them as well.

For the further data reduction we used the raw COS spectra 
of the individual science exposures. The individual exposures 
were processed with the CALCOS pipeline (v2.17.3) to produce 
the standard COS {\tt x1d} fits files. For the coaddition of the individual 
spectra we then used a custom-written code that aligns
the individual exposures in wavelength space in a fully automated fashion. 
The code calculates for each exposure
a pixel/wavelength calibration based on the line flanks (for
spectra with S/N$>5$) or line centers (for spectra with S/N$\leq 5$) 
of various interstellar anchor lines that are distributed over the 
wavelength range of the G130M and G160M spectral ranges.
The heliocentric velocity positions of the anchor lines were
calibrated for each sightline using H\,{\sc i} 21cm data from the 
Leiden-Argentine-Bonn (LAB) survey (Kalberla et al.\,2005). 
The individual spectra then were uniformly rebinned and coadded pixel-by-pixel
(using the count rate in each pixel), where pixels with
known artifacts were flagged accordingly. Errors were calculated
in the coadded spectra by weighting by inverse variance.
In this way, we obtained for each sightline a calibrated, co-added 
G130M spectrum (and G160M spectrum, if available). For each 
sightline we checked the quality of the data reduction by 
a visual inspection of the final spectrum.

As it turns out, many of the data sets have very low signal-to-noise
ratios (S/N) and/or sample background sources with very low redshifts. 
These spectra are thus not suited to investigating intervening
metal absorption at $z\leq 0.1$. We selected only those spectral data 
that have a minimum S/N per resolution element of four in 
the wavelength range between $1208$ and $1338$ \AA\, and for which the redshift
of the background source is $z_{\rm QSO}>0.03$. This 
selection reduces the total sample 
to $303$ lines of sight (LOS). In Tables A.1-A.4 in the Appendix we
present a complete list of all 303 QSO sightlines in our COS sample 
including QSO names and coordinates.

\subsection{Redshift-path coverage, absorber identification, and spectral analysis}

The next step in our analysis was to characterize the available redshift path to
detect intervening Si\,{\sc iii}+H\,{\sc i}\,Ly\,$\alpha$ absorption at $z\leq0.1$ 
in each spectrum. 
We generally exclude the velocity range between $|v_{\rm helio}|=0-500$
km\,s$^{-1}$ where absorption by local disk gas and by intermediate- and high-velocity
clouds in the Milky Way halo is found (see, e.g, Wakker \& van Woerden 1998;
Richter 2006; Shull et al.\,2009; Lehner et al.\,2012; Putman, Peek \& Joung 2012;
Herenz et al.\,2013).
To sort out regions that may be associated with the background AGN we further 
ignore the spectral range that lies within $5000$ km\,s$^{-1}$ of $z_{\rm QSO}$.
Finally, we identify and flag along each sightline those spectral regions that
are heavily blended by other intervening absorbers (in particular by higher-redshift 
LLS and DLAs). 
As a result, we obtain a total absorption redshift path $\Delta z_{\rm abs}<0.1$ 
for each sightline that is available to identify intervening 
Si\,{\sc iii}+H\,{\sc i}\,Ly\,$\alpha$ absorption (see Tables A.1-A.4, last column).

For the determination of the Si\,{\sc iii} number density (Sect.\,4.2) we further need 
to consider the detection limit for Si\,{\sc iii} absorption along each sightline,
which depends on the S/N in the relevant part of the spectrum where 
intervening Si\,{\sc iii} absorption is expected to occur.
The minimum column density, $N_{\rm lim}$, that can be detected at $3\sigma$ significance 
from an unresolved absorption line with a laboratory wavelength $\lambda_0$ and
an oscillator strength $f$ in a spectrum with a resolving power $R$ and a 
given S/N per resolution element is given by 
(e.g., Richter et al.\,2001; Tumlinson et al. 2002):

%%%%%%%%%%%%%%%%%%%%%%%%%%%%%%%%%%%%%%%%%%%%%%%%%%%%%%%%

\begin{equation}
N_{\rm lim}\approx1.13\times10^{20}\,
\frac{3}{R\,{\rm (S/N)}\,f\,(\lambda_0 /{\rm A})}\,{\rm cm}^{-2}.
\end{equation}

%%%%%%%%%%%%%%%%%%%%%%%%%%%%%%%%%%%%%%%%%%%%%%%%%%%%%%%%

For the only available Si\,{\sc iii} transition in the UV range we 
have $\lambda_0=1206.500$ \AA\, and $f=1.62732$ (Morton 2003), so that for 
a S/N of $4$ per resolution element the formal $3\sigma$ detection 
limit for $R=15,000$ is log $N_{\rm lim}$(Si\,{\sc iii}$)=12.46$.
More than half of the considered COS spectra have a S/N per 
resolution element of $\geq 12$, so that for these LOS 
log $N_{\rm lim}$(Si\,{\sc iii}$)\leq12.0$. 

For the identification of intervening Si\,{\sc iii} absorbers at $z\leq0.1$ in
our COS data sample we used the following strategy. In a first step,
we let an automated line-finder algorithm identify absorption features whose 
wavelengths would correspond to combined Si\,{\sc iii}/H\,{\sc i}\,Ly\,$\alpha$
absorption and created a candidate list of possible Si\,{\sc iii} absorbers.
In a second step, we inspected each individual spectrum by eye and
created a different (independent) candidate list. For each of the candidates from both
lists velocity plots (including all relevant low and high ions in the available
wavelength range) were generated and the absorber candidates were analyzed in detail.
To meet our selection criteria of a bona fide 
Si\,{\sc iii} absorber, we require the significant
detection of Si\,{\sc iii} and H\,{\sc i} {\it together} with the detection 
of at least one additional metal ion (C\,{\sc iv}, C\,{\sc ii}, Si\,{\sc iv}, Si\,{\sc ii})
to avoid mis-indentifications and false detections due to absorption features of 
other intervening absorbers at higher redshifts. 
Si\,{\sc iii} traces both diffuse ionized gas (as also traced by high ions 
such as C\,{\sc iv} and Si\,{\sc iv}) as well as denser, partly neutral gas 
(as also traced by singly-ionized species such as Si\,{\sc ii} and Mg\,{\sc ii};
see Fig.\,1), so that the simultaneous detection of Si\,{\sc iii} with 
either one of the above listed low and high ions is expected.

We also identified a number of absorber candidates, that possibly show 
absorption in H\,{\sc i} and Si\,{\sc iii} but lack absorption from
any other ion, so that a unambiguous identification cannot be given.

%%%%%%%%%%%%%%%%%%%%%% FIGURE 03 %%%%%%%%%%%%%%%%%%%%%%

\begin{figure}[t!]
\resizebox{1.0\hsize}{!}{\includegraphics{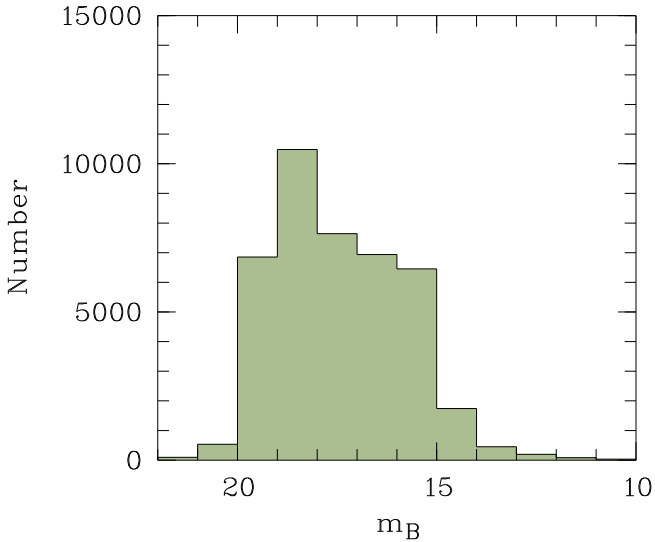}}
\caption[galaxy sample]{
Distribution of $B$ band magnitudes for $\sim 41,000$ galaxies in
our sample.
}
\end{figure}

%%%%%%%%%%%%%%%%%%%%%%%%%%%%%%%%%%%%%%%%%%%%%%%%%%%%%%%%

For each intervening Si\,{\sc iii} absorber we checked for associated 
absorption in other metal ions (in particular Si\,{\sc ii}, Si\,{\sc iv},
C\,{\sc ii}, C\,{\sc iv}, N\,{\sc v}, Fe\,{\sc ii}).
We then measured equivalent widths (and their limits) for the strongest lines 
of the three different ionization states of silicon 
(Si\,{\sc ii} $\lambda 1260$, Si\,{\sc iii} $\lambda 1206$, Si\,{\sc iv} $\lambda 1393$),
C\,{\sc ii} $\lambda 1334$, and C\,{\sc iv} $\lambda 1548$, 
as well as for H\,{\sc i}\,Ly\,$\alpha$ by a direct pixel integration.
Because of the limited spectral resolution and S/N of the COS data and the resulting 
lack of information about the intrinsic component structure in the absorbers 
we refrain from performing line fits for the analysis of the metal absorption in the systems. 
We instead use the apparent-optical depth method (AOD method; Savage \& Sembach 1991) 
to derive column densities for unsaturated absorption in the above-listed lines and 
lower limits for lines that appear to be saturated. We assume that saturation becomes 
important for lines that have an absorption depth $>0.3$ at the resolution of COS. 
Only for a small number of absorbers do we use Voigt-profile fitting to estimate 
the total H\,{\sc i} column density from the damping wings of the H\,{\sc i} 
Ly\,$\alpha$ absorption. The AOD method and Voigt-profile fitting are implemented
in the custom-written {\tt span} software package that is based on ESO-MIDAS
and its data-reduction libraries (Fontana \& Ballester 1995).
Laboratory wavelengths and oscillator strengths for all ions were adopted 
from the compilation by Morton (2003). 

%%%%%%%%%%%%%%%%%%%%%% FIGURE 04 %%%%%%%%%%%%%%%%%%%%%%

\begin{figure*}[ht!]
\resizebox{1.0\hsize}{!}{\includegraphics{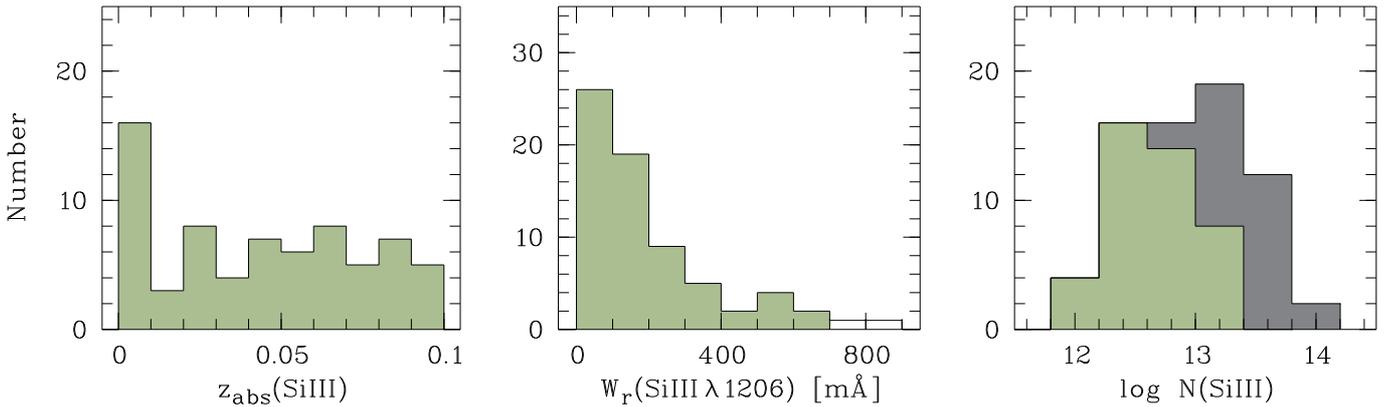}}
\caption[Absorption properties SiIII]{
{\it Left panel:} Redshift distribution of the 69 intervening
Si\,{\sc iii} absorbers in our sample.
{\it Middle panel:} Distribution of Si\,{\sc iii} $\lambda 1206$ restframe
equivalent widths of the 69 absorbers, obtained from the pixel integration
over the Si\,{\sc iii} velocity profiles.
{\it Right panel:} Distribution of logarithmic Si\,{\sc iii} column densities
for the 69 absorbers, derived using the AOD method. The gray-shaded area indicates
lower limits for log $N$(Si\,{\sc iii}) for absorbers that have saturated
Si\,{\sc iii} $\lambda 1206$ lines.
}
\end{figure*}

%%%%%%%%%%%%%%%%%%%%%%%%%%%%%%%%%%%%%%%%%%%%%%%%%%%%%%%%

\subsection{Galaxy data}

To statistically investigate the relation between Si\,{\sc iii} absorbers and 
low-redshift galaxies we searched for publicly available galaxy data in 
the SIMBAD data archive
\footnote{\tt http://simbad.u-strasbg.fr}.
Using SIMBAD we generated a list of galaxies with known redshifts located 
at $z\leq 0.1$ and within $2\deg$ of each sightline. A redshift of $z=0.1$ 
corresponds to a proper distance of $420.9$ Mpc at $z=0.1$ for a standard 
$\Lambda$CDM cosmology with $\Omega_{\Lambda}=0.72, \Omega_{\rm m}=0.28$ and
$H_0=69.7$ km\,s$^{-1}$\,Mpc$^{-1}$ (Hinshaw et al.\,2013
\footnote{These values are consistent at the 2$\sigma$ level with the
latest measured values by the Planck Collaboration (Planck Collaboration 2015. {\sc xiii}).}).
In this way, we obtained redshifts and coordinates for $64,280$ galaxies distributed 
around the $303$ COS sightlines. 
For $40,907$ of these galaxies we have additional information on their $B$-band 
magnitudes. The distribution of $m_B$ for these systems is shown in Fig.\,3. 
The distribution breaks down at $m_B=20.0$, with only two percent of the galaxies 
having $m_B>20.0$. For $z=0.1$ and the above given cosmology, $m_B=20.0$ 
corresponds to an absolute magnitude of $M_B\approx -18.1$. If we consider the 
$g$-band SDSS galaxy-luminosity function (Montero-Dorta \& Prada 2009) as reference, 
this value translates into a lower luminosity cutoff of $L\approx 0.5\,L^{\star}$.

For each sightline, we then calculated the projected impact parameters, $\rho$, of 
the galaxies to the LOS. There are $11,127$ galaxies within $\rho=1$ Mpc located 
in the cylindrical volume around the $303$ sightlines. This number implies a mean 
galaxy density of $\phi\approx0.03$ Mpc$^{-3}$ in our total galaxy sample, which 
is about seven times the space density of $L^{\star}$ galaxies in SDSS $g$-band 
(Montero-Dorta \& Prada 2009).

At this point it is important to mention that the galaxy catalog created in this 
way is highly inhomogeneous because the data stems from different galaxy surveys and
pointed observations. As a result, the individual values for $\phi$ for each sightline 
vary considerably. In terms of luminosities, only a few sightlines have galaxy data 
that are sufficiently deep to detect faint ($L\leq 0.05\,L^{\star}$) galaxies. In
general, our galaxy data is incomplete for $L<0.5L^{\star}$. This aspect needs to 
be carefully taken into account for the interpretation of the observed absorber/galaxy 
relation. For many sightlines, we are missing low-luminosity galaxies that
may be close to the LOS and/or responsible for the observed metal absorption
(see discussion in Sect.\,6).
We do not consider any other morphological parameters of the
selected galaxies (Hubble type, star-formation rate, etc.), because such 
information is available to us only for a small fraction of the galaxies 
in our sample.

%%%%%%%%%%%%%%%%%%%%%%%%%%%%%%%%%%%%%%%%%%%%%%%%%%%%%%%%

\section{Absorber properties}

\subsection{Si\,{\sc iii} detection rate, equivalent widths and column densities}

Using the above outlined strategy, we identify 69 intervening Si\,{\sc iii} 
absorbers along the 303 selected QSO sightlines. 
The statistical and physical
properties of these 69 intervening absorbers are discussed in this section.

In Tables A.5-A.10 in the Appendix we list the measured equivalent widths
and derived column densities for the various ions detected in these systems.
These tables also contain information
on two additional strong metal absorbers towards SDSSJ141542.90+163413.8
and UKS$-$0242-724 that are detected
in Si\,{\sc ii}, Si\,{\sc iv}, and other ions, but not in Si\,{\sc iii},
because there are no useful data in the relevant spectral region where
Si\,{\sc iii} is expected. 
In Fig.\,2 we show five examples for velocity profiles of 
intervening Si\,{\sc iii} absorbers with different
Si\,{\sc iii} absorption strengths. The complete set of 
velocity profiles for all 69 confirmed Si\,{\sc iii} absorbers is provided
in Figs.\,A.1-A.7 in the Appendix of this paper.
Following the criteria defined in the previous section we further 
identify 20 Si\,{\sc iii} candidate absorbers along the 303 lines
of sight. Although these candidate systems are not considered
in the further analysis, we list the QSO names and (possible) absorption 
redshifts of these systems in Table A.11 in the Appendix.

The redshift distribution of the 69 absorbers is shown in Fig.\,4, left
panel. The absorption redshifts lie between $z=0.00014$ and $z=0.09789$.
While for $z=0.01-0.10$ the absorbers are uniformly distributed over 
the surveyed redshift range, the distribution shows a peak
in the first redshift bin at $z<0.01$. This peak is related
to an overdensity of Si\,{\sc iii} absorbers in the broader Virgo-Cluster 
environment that is traced along several sightlines in our QSO sample.
In the middle panel of Fig.\,4 we show the distribution of Si\,{\sc iii} 
$\lambda 1206$ equivalent widths of the 69 absorbers. The distribution
peaks at relatively low equivalent widths ($W_{\rm r}\leq 100$ m\AA);
65 percent of the absorbers have $W_{\rm r}\leq 200$ m\AA, while
most of the remaining 35 percent are spread over a large range in $W_{\rm r}$
between 200 and 700 m\AA.
There are two systems that have very large equivalent widths of 
$W_{\rm r}>700$ m\AA, belonging to strong absorption systems 
towards SDSSJ140732.25+550725.6 and PG\,0832+251 (see Appendix).

Si\,{\sc iii} column densities for the 69 absorbers, as derived from the 
AOD method (see previous section), are shown in the right panel of Fig.\,4.
The gray-shaded area (which adds to the green-shaded area) 
indicates lower limits of $N$(Si\,{\sc iii}) for absorbers
where the Si\,{\sc iii} $\lambda 1206$ line shows evidence for saturation.
The decline of the observed distribution at low column densities 
reflects both the inhomogeneous S/N in the spectra as well as the 
column density distribution that is intrinsic to the absorber population.
From the estimate of $N_{\rm lim}$(Si\,{\sc iii}) for each sightline 
(see equation 1) follows that more than 90 percent of all sightlines
are sensitive to detect Si\,{\sc iii} absorbers with log $N$(Si\,{\sc iii}$)<12.2$,
but only four such systems are found in our data 
(fractional abundance $4/69\approx0.04$). Therefore, intervening
Si\,{\sc iii} absorbers with log $N$(Si\,{\sc iii}$)<12.2$ are
rare and may even represent a population that is distinct from
the absorbers with higher column densities (as is discussed later).
At the high-column density end, the distribution breaks down at 
log $N$(Si\,{\sc iii}$)=14$. Even if some of the saturated absorbers 
(gray-shaded range) would have logarithmic Si\,{\sc iii} column densities  
$>14$, their number would be small compared to absorbers with 
log $N$(Si\,{\sc iii}$)<14$. Our conclusion is that the
{\it characteristic} column density range for intervening Si\,{\sc iii}
absorbers is log $N$(Si\,{\sc iii}$)=12.2-14.0$.

\subsection{Number density of Si\,{\sc iii} absorbers}

For the determination of the number density of intervening
Si\,{\sc iii} absorbers per unit redshifz, $d{\cal N}/dz$,
we need to consider in detail the completeness of our
absorber survey and the selection bias in our QSO sample.

As discussed above, log $N_{\rm lim}$(Si\,{\sc iii}$)\geq 12.2$
represents the column density range that is characteristic for intervening
Si\,{\sc iii} absorbers. 63 Si\,{\sc iii} absorbers in our sample have column
densities log $N_{\rm lim}$(Si\,{\sc iii}$)\geq 12.2$ and 280 out of the 303 
sightlines are sensitive to this level, covering a total redshift path
of $\Delta z_{\rm abs}=21.95$. The resulting number density thus is
$d{\cal N}/dz=2.9\pm 0.4$. The given errors represent statistical
errors calculated from Poisson statistics.
For log $N_{\rm lim}$(Si\,{\sc iii}$)=11.8$ we derive $d{\cal N}/dz=4.0\pm 0.7$
(31 absorbers, $\Delta z_{\rm abs,tot}=7.8$), 
while for log $N_{\rm lim}$(Si\,{\sc iii}$)=12.6$
the number density is slightly smaller ($d{\cal N}/dz=2.1\pm 0.3$, 49 absorbers,
$\Delta z_{\rm abs,tot}=23.6$). 
If we consider only low-column density absorbers with 
log $N$(Si\,{\sc iii}$)=11.8-12.2$
we obtain $d{\cal N}/dz\approx0.3$.

One important effect that influences the observed frequency of intervening
absorption systems is the selection bias in the QSO sample that we are using. 
Most of the bright QSOs in our COS sample have
been previously studied using instruments such as \emph{HST}/STIS,
\emph{FUSE}, and others. While the original motivation to select these
background sources certainly was their intrinsic brightness, the re-observation
of these sources with COS and the chosen integration time (e.g., the achieved
S/N), possibly was motivated by targeting particularly interesting
intervening absorbers to perform a detailed study of these systems.
Therefore, one of the reasons for limiting the absorber sample to redshifts
$z\leq0.1$ was to avoid the inclusion of targeted observations of stronger
intervening metal absorbers at $z>0.12$ that can be observed in both
H\,{\sc i} Ly\,$\alpha$ as well as Ly\,$\beta$.

We have scanned the various original COS proposals that outline the motivation
for observing the LOS along which we detect intervening Si\,{\sc iii} at $z\leq0.1$.
As it turns out, our QSO sample covers most of the sightlines 
selected for the COS-Dwarfs survey (Bordoloi et al.\,2014), a targeted survey of
sightlines passing through the virial radius of $z\leq 0.1$ dwarf galaxies to
study the CGM of these systems. For an un-biased estimate of 
$d{\cal N}/dz$(Si\,{\sc iii}) all these sightlines need to be excluded.
All in all, we identify 40 QSO sightlines in our sample that have been 
specifically observed to study circumgalactic gas in the vicinity of known
low-redshift galaxies. 
By removing these sightlines from our sample we derive the following
bias-corrected number densities of Si\,{\sc iii} absorbers for the 
different limiting Si\,{\sc iii} column densities: 
$d{\cal N}/dz=3.8\pm 0.7$ for log $N_{\rm lim}$(Si\,{\sc iii}$)=11.8$,
$d{\cal N}/dz=2.5\pm 0.4$ for log $N_{\rm lim}$(Si\,{\sc iii}$)=12.2$, and
$d{\cal N}/dz=1.7\pm 0.3$ for log $N_{\rm lim}$(Si\,{\sc iii}$)=12.6$.
For the range log $N$(Si\,{\sc iii}$)=11.8-12.2$ we again obtain 
$d{\cal N}/dz\approx0.3$.

It is evident that these numbers are not substantially different from
those derived from the biased sample. This is, however, not surprising
because it is known that the UV absorption cross section of the warm CGM 
around {\it dwarf galaxies} appears to be small when compared to the warm CGM 
of more massive galaxies (e.g., Bordoloi et al.\,2014; Liang \& Chen 2014;
see also Sect.\,8.3).
In other words: only a few {\it additional} Si\,{\sc iii} 
absorbers are (in a statistical sense) added to our absorber sample 
when including the pre-selected Bordoloi et al. sightlines,
but because of the overall large size of our QSO sample their 
influence on $d{\cal N}/dz$ is small.

In Fig.\,5 we show the bias-corrected number density of intervening 
Si\,{\sc iii} absorbers per unit redshift in our survey, $d{\cal N}/dz$, for the 
three different limiting Si\,{\sc iii} column densities, 
log $N_{\rm lim}$(Si\,{\sc iii}$)=11.8,12.2$ (dashed red line),
and $12.6$. In the following, we refer to these values when discussing 
the number densities of intervening Si\,{\sc iii} absorbers.

%%%%%%%%%%%%%%%%%%%%%% FIGURE 05 %%%%%%%%%%%%%%%%%%%%%%

\begin{figure}[t!]
\begin{center}
\resizebox{0.8\hsize}{!}{\includegraphics{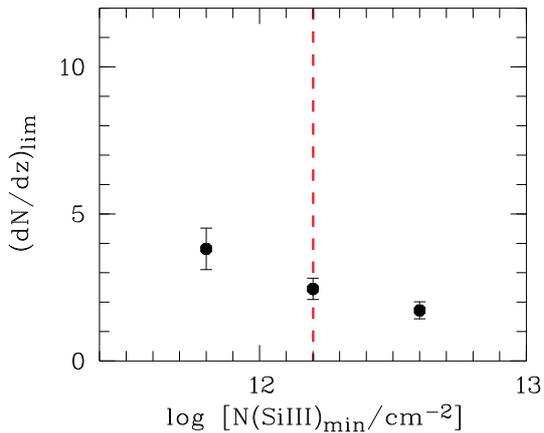}}
\caption[SiIII number densities]{
Number densities, $d{\cal N}/dz$, of intervening Si\,{\sc iii} absorbers for
different limiting column densities.
}
\end{center}
\end{figure}

%%%%%%%%%%%%%%%%%%%%%% FIGURE 06 %%%%%%%%%%%%%%%%%%%%%%

\begin{figure*}[ht!]
\resizebox{1.0\hsize}{!}{\includegraphics{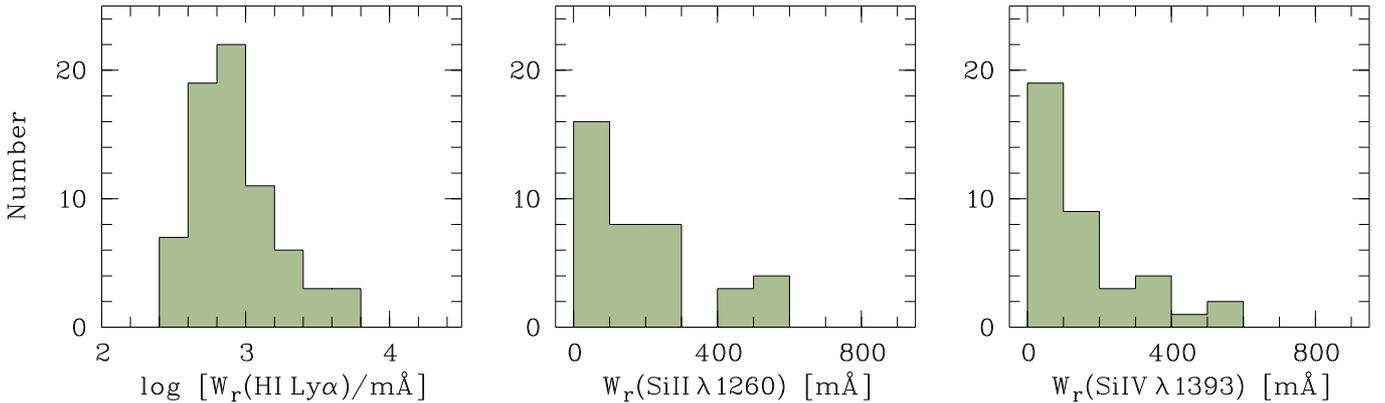}}
\caption[Absorption properties SiIII]{
Distribution of restframe equivalent widths of H\,{\sc i}\,Ly\,$\alpha$
({\it left panel}), Si\,{\sc ii} $\lambda 1260$ ({\it middle panel}),
and Si\,{\sc iv} $\lambda 1393$ ({\it right panel}) associated with
intervening Si\,{\sc iii} absorbers.
}
\end{figure*}

%%%%%%%%%%%%%%%%%%%%%%%%%%%%%%%%%%%%%%%%%%%%%%%%%%%%%%%%

\subsection{Associated absorption from other ions}

\subsubsection{H\,{\sc i}}

As outlined in the introduction, the Si\,{\sc iii} ion is,
because of its ionization potential, a very sensitive tracer for
both predominantly neutral as well as predominantly ionized gas in the
halos of galaxies. As a consequence, the
H\,{\sc i} column density in the Si\,{\sc iii} absorbing gas 
varies over several orders of magnitudes, ranging from 
log $N$(H\,{\sc i}$)\sim 14$ up to log $N$(H\,{\sc i}$)\sim 22$,
depending on whether the sight line passes through a galaxy disk 
or a galaxy halo, and on the local ionization conditions.
Detailed ionization models of the Si\,{\sc iii} absorbers
are presented in Sect.\,7.
In the left panel of Fig.\,6 we show the distribution of
H\,{\sc i} Ly\,$\alpha$ equivalent widths in the Si\,{\sc iii}-selected 
absorbers, which similarly span an extremely large range 
from $W_{\rm r}\approx 250$ m\AA\, to $W_{\rm r}\approx 5$
\AA. We generally refrain from estimating the H\,{\sc i} column densities
from the saturated H\,{\sc i} Ly\,$\alpha$ lines because of the very large
uncertainties that such an estimate would be afflicted with. 

It is worth noting that our absorber sample contains four 
damped Ly\,$\alpha$/sub-damped Ly\,$\alpha$ systems (DLAs/sub-DLAs) with
log $N$(H\,{\sc i}$)>19.2$, as estimated from fitting the damping wings
of the Ly\,$\alpha$ absorption in these systems. With a 
total redshift path of $\Delta z_{\rm abs,tot}=21.95$, this absorber frequency corresponds
to a number density of $d{\cal N}/dz\approx0.2$. 
Despite the low-number statistics, it is worth mentioning this value agrees 
well with the expected number density of sub-DLAs/DLAs as estimated 
from the H\,{\sc i} mass function of $z\approx0$ galaxies (Zwaan et al.\,2005).

\subsubsection{Si\,{\sc ii}} 

Detailed ionization models (see Sect.\,7) indicate that regions with gas densities 
log $n_{\rm H}\geq -3.5$ and sufficiently large neutral hydrogen
column densities are expected to show both Si\,{\sc iii} as well as
Si\,{\sc ii} absorption. Because Si\,{\sc ii} and Mg\,{\sc ii}
have almost identical ionization potentials (Morton 2003) and silicon
and magnesium have similar solar abundances (Asplund et al.\,2009), the 
absorber statistics for Si\,{\sc ii} can be directly compared with
the well-established statistics on intervening Mg\,{\sc ii} 
absorbers.
The presence of Si\,{\sc ii} absorption {\it without} associated 
Si\,{\sc iii} is basically impossible for any realistic 
galactic or circumgalactic gas environment (see Sect.\,7), so that
the fact that our absorber sample is Si\,{\sc iii}-selected is not
expected to introduce a selection bias in our Si\,{\sc ii} statistics.

Fourty of our Si\,{\sc iii} absorbers show associated Si\,{\sc ii} absorption
in the strongest of the available Si\,{\sc ii} lines at $\lambda_0=1260.42$ \AA.
The equivalent-width distribution for Si\,{\sc ii} $\lambda 1260$ in 
these absorbers is shown in the middle panel of Fig.\,6. It
shows that the majority (80 percent) of the detected Si\,{\sc ii} lines 
have equivalent widths in the $\lambda 1260$ line of $<300$ m\AA.
To compare the observed number density of Si\,{\sc ii}-bearing absorbers with that
of the prominent strong Mg\,{\sc ii} systems, i.e., systems that have 
an equivalent width of $W_{\rm r}\geq 300$ m\AA\,in the Mg\,{\sc ii}
$\lambda 2976$ line, we need to convert the observed Si\,{\sc ii} $\lambda 1260$
equivalent widths into Mg\,{\sc ii} $\lambda 2976$ equivalent widths.
For this we use the Si\,{\sc ii}/Mg\,{\sc ii} conversion scheme presented in 
Herenz et al.\,(2013), after which an equivalent widths of $W_{\rm r}=300$ m\AA\,
in the Mg\,{\sc ii} $\lambda 2976$ line corresponds to an equivalent width of 
$W_{\rm r}=140$ m\AA\,in Si\,{\sc ii} $\lambda 1260$, assuming solar 
relative abundances of Mg and Si. In our absorber sample we have 22
systems that have $W_{\rm r}\geq140$ m\AA\,in the Si\,{\sc ii} $\lambda 1260$
line, from which we indirectly infer a number density of strong Mg\,{\sc ii} absorbers
at $z\leq0.1$ of $d{\cal N}/dz\approx 1.0$. This value is in good agreement
with the expectations for $d{\cal N}/dz$(Mg\,{\sc ii} $\lambda 2976$) based on the
redshift evolution of strong Mg\,{\sc ii} absorbers in SDSS data ($d{\cal N}/dz \sim 0.8$;
Nestor, Turnshek \& Rao 2005; Prochter et al.\,2006; Lundgren et al.\,2009;
Zhu \& Menard 2013).

\subsubsection{Si\,{\sc iv}}

There are 38 Si\,{\sc iii} absorbers in our sample that show associated 
Si\,{\sc iv} absorption (the two available Si\,{\sc iv} transitions are located
at $\lambda_0=1393.8$ and $1402.8$ \AA, the former being the stronger
of the two transitions; Morton 2003). The distribution of Si\,{\sc iv} 
$\lambda 1393$ equivalent widths of the 38 absorbers detected in
Si\,{\sc iv} is shown in the right panel of Fig.\,6. Similarly as for
Si\,{\sc ii}, the majority of the Si\,{\sc iv} $\lambda 1393$ equivalent widths 
are small (82 percent have $W_{\rm r}(1393)<300$ m\AA). 
From the observed frequency of Si\,{\sc iv} absorption
in the Si\,{\sc iii}-selected systems we estimate 
$d{\cal N}/dz$(Si\,{\sc iv})$)\geq 1.5$ for 
log $N$(Si\,{\sc iv}$)\geq 12.6$. The occurrence
of Si\,{\sc iv} absorption {\it without} associated Si\,{\sc iii} absorption
in the CGM/IGM is possible, in principle, but is relevant only for absorbers 
at relatively low gas densities (log $n_{\rm H}\leq -4.2$; see Sect.\,7).
Still, we formally can only give a lower limit for $d{\cal N}/dz$(Si\,{\sc iv})
from our survey.

%%%%%%%%%%%%%%%%%%%%%% FIGURE 07 %%%%%%%%%%%%%%%%%%%%%%

\begin{figure*}[ht!]
\begin{center}
\resizebox{1.0\hsize}{!}{\includegraphics{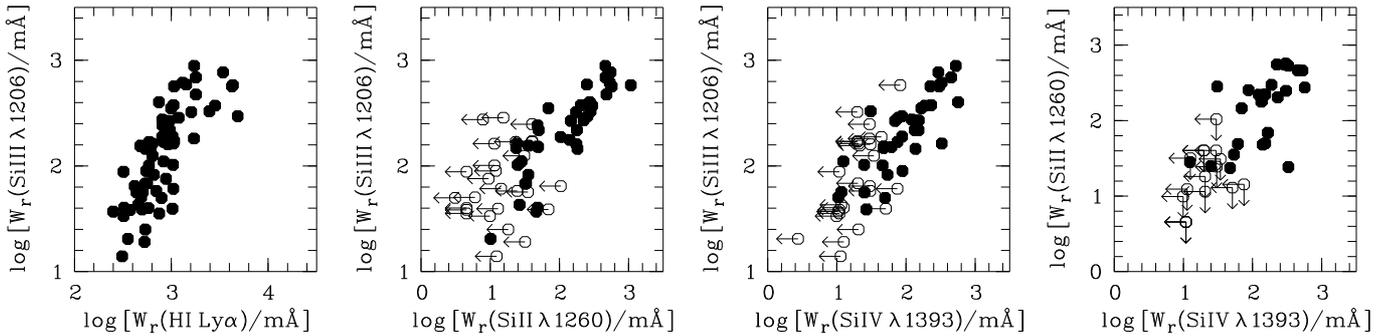}}
\caption[Equivalent widths II]{
Comparison between restframe equivalent widths of Si\,{\sc iii}, Si\,{\sc ii},
Si\,{\sc iv}, and H\,{\sc i} for 69 absorbers.
{\it First panel (left):} Si\,{\sc iii} $\lambda 1206$ vs.\,H\,{\sc i} Ly\,$\alpha$;
{\it Second panel:} Si\,{\sc iii} $\lambda 1206$ vs.\,Si\,{\sc ii} $\lambda 1260$;
{\it Third panel:} Si\,{\sc iii} $\lambda 1206$ vs.\,Si\,{\sc iv} $\lambda 1393$;
{\it Fourth panel:} Si\,{\sc ii} $\lambda 1260$ vs.\,Si\,{\sc iv} $\lambda 1393$.
}
\end{center}
\end{figure*}

%%%%%%%%%%%%%%%%%%%%%%%%%%%%%%%%%%%%%%%%%%%%%%%%%%%%%%%%

\subsection{Correlation plots}

In Fig.\,7 we show correlations between the measured (logarithmic) 
equivalent widths of Si\,{\sc iii} $\lambda 1206$, H\,{\sc i} Ly\,$\alpha$,
Si\,{\sc ii} $\lambda 1260$, and Si\,{\sc iv} $\lambda 1393$ for absorbers
in which the absorption in these ions is aligned in velocity space
within one COS resolution element ($20$ km\,s$^{-1}$). 
The interpretation of the observed correlations involves the possible
presence of different gas phases in the absorbers that may or may not be
co-spatial within the overall gas structures.\\

\subsubsection{Si\,{\sc iii} vs.\,H\,{\sc i}} 

For log $[W_{\rm r}$(Ly\,$\alpha)]\leq 3$ 
the Si\,{\sc iii} $\lambda 1206$ equivalent width rises steeply with the mildly 
increasing equivalent width of the fully saturated H\,{\sc i} Ly\,$\alpha$ absorption
(Fig.\,7; left panel).
This trend indicates (as expected) that Si\,{\sc iii} and H\,{\sc i} trace the same physical
regions that span a large range in neutral (and total) gas column densities.
For log $[W_{\rm r}$(Ly\,$\alpha)]>3$ the correlation turns over into a somewhat flatter
regime because the Si\,{\sc iii} $\lambda 1206$ absorption itself becomes 
saturated at such high total gas columns.

\subsubsection{Si\,{\sc iii} vs.\,Si\,{\sc ii}}

The equivalent widths of Si\,{\sc iii} $\lambda 1206$ and Si\,{\sc ii} $\lambda 1260$ 
clearly are correlated with each other in those absorbers, where both ions are detected
(Fig.\,7; second panel from left, filled circles). 
This demonstrates that part of the Si\,{\sc iii} arises in the same gas 
phase as Si\,{\sc ii}, i.e., in predominantly neutral and/or mildly ionized 
gas. Some of the systems with upper limits in $W_{\rm r}(\lambda 1260)$ (open
circles/arrows) have relatively strong Si\,{\sc iii} absorption without
a Si\,{\sc ii} counterpart, implying that these absorbers consist of 
predominantly ionized gas that is traced by higher ions (e.g., Si\,{\sc iv},
see below).

\subsubsection{Si\,{\sc iii} vs.\,Si\,{\sc iv}}

For systems that show both Si\,{\sc iii} as well as Si\,{\sc iv} absorption
the equivalent width of Si\,{\sc iii} $\lambda 1206$ is also correlated with that
of Si\,{\sc iv} $\lambda 1393$, as can be seen in the third panel
of Fig.\,7 (filled circles). This trend indicates that some part of the 
Si\,{\sc iii} resides in a more ionized gas phase that is traced by 
Si\,{\sc iv} (and other high ions such as C\,{\sc iv}). The relation is mildly
steeper than the one for Si\,{\sc iii}/Si\,{\sc ii}, but has
as similar level of scatter.

\subsubsection{Si\,{\sc ii} vs.\,Si\,{\sc iv}}

In the right panel of Fig.\,7 we have plotted the equivalent width
of Si\,{\sc ii} $\lambda 1260$ against that of Si\,{\sc iv} $\lambda 1393$ 
for the absorbers where both ions are detected. The total number of systems that show 
aligned Si\,{\sc ii} and Si\,{\sc iv} absorption is relatively small
($38$ percent). This implies 
that both ions predominantly trace complementary gas phases. 
For the systems detected in both ions (filled circles) 
the correlation between  $W_{\rm r}$(Si\,{\sc ii} $\lambda 1260$) and 
$W_{\rm r}$(Si\,{\sc iv} $\lambda 1393$) is weak and
shows a relatively large scatter.

\subsubsection{Interpretation}

We conclude that the correlation plots between the equivalent widths of 
Si\,{\sc iii} $\lambda 1206$, H\,{\sc i} Ly\,$\alpha$,
Si\,{\sc ii} $\lambda 1260$, and Si\,{\sc iv} $\lambda 1393$ are in
line with (and further support) the idea that the Si\,{\sc iii} absorption 
in intervening metal-systems traces metal-enriched gas within a wide range 
of physical conditions including a) a denser (partly neutral) phase also 
traced by Si\,{\sc ii} and other low ions, and b) a more diffuse (predominantly
ionized) gas phase also traced by Si\,{\sc iv} and other high ions. 
We further discuss these aspects in Sect.\,7 where we model the 
ionization conditions in Si\,{\sc iii}/Si\,{\sc iv} absorbers in detail.

In a recent C\,{\sc iv}-selected survey of absorbers
at $z\leq 0.16$ Burchett et al.\,(2015) find very similar relations 
between Si\,{\sc ii}, Si\,{\sc iii}, and Si\,{\sc iv} in their 
absorber sample (their Figs.\,15 and 16).

%%%%%%%%%%%%%%%%%%%%%%%%%%%%%%%%%%%%%%%%%%%%%%%%%%%%%%%%

\section{On the expected cross section of metal-enriched gas in galaxy halos}

Before we investigate in detail the {\it observed} relation between
Si\,{\sc iii} absorbers and galaxies in a statistical sense, we first set up
a geometrical model to estimate the expected cross section of metal-enriched
gas in the extended gaseous halos of galaxies at $z=0$.

Under the assumption that {\it all} intervening Si\,{\sc iii} absorbers 
at low $z$ are related to metal-enriched gas situated 
in the extended halos of galaxies, the observed number density of Si\,{\sc iii} 
absorbers can be directly linked to the space density of galaxies, $\phi$, and 
the effective (geometrical) cross section of the absorbing gas, which is a 
product of the projected area covered by the gaseous halo $A_{\rm halo}=\pi r_{\rm halo}^2$
and the mean covering fraction $\langle f_{\rm c}\rangle$ of the gas phase
that is seen in absorption:

%%%%%%%%%%%%%%%%%%%%%%%%%%%%%%%%%%%%%%%%%%%%%%%%%%%%%%%%

\begin{equation}
\frac{d{\cal N}}{dz}=
\phi\,\langle f_{\rm c}\rangle\,A_{\rm halo}\,\frac{c\,(1+z)^2}{H(z)}.
\end{equation}

%%%%%%%%%%%%%%%%%%%%%%%%%%%%%%%%%%%%%%%%%%%%%%%%%%%%%%%%

The Hubble parameter is defined as 
$H(z)=H_0\,(\Omega_{\rm m}\,(1+z)^3+\Omega_{\Lambda})^{1/2}$, which is 
appropriate for a matter-dominated flat Universe with a 
cosmological constant.
Equation (2), and modified versions of it, have been commonly used to 
estimate the sizes of galaxy halos and the covering fractions of individual
ions from QSO absorption-line observations
(e.g., Kapzcrak et al.\,2008; Richter et al.\,2011; Prochaska et al.\,2011).

We here take the opposite point of view and pose the following 
question: What would be the expected number density of intervening 
metal absorbers, if all galaxies at $z=0$ contain detectable 
metal-enriched gas in their halos that extends exactly to their 
respective virial radii?

Using equation (2) it is indeed relatively straight-forward to set up a "toy model" for the 
absorption cross section of extended halo gas taking into account the observed space density 
and luminosity/mass distribution of galaxies at low redshift.
The motivation for such a reverse approach is rather simple: {\it if we could know the 
maximum contribution of metal-enriched gas that is gravitationally bound to
galaxies to the number density of intervening metal absorbers, we would have an
important reference value for the interpretation of the observed number densities
of intervening} Si\,{\sc iii} {\it systems and their origin in the CGM and/or IGM}.

Both the galaxy density $\phi$ as well as the distribution of the galaxies' 
virial radii at $z=0$, i.e., the most important parameters 
to calculate $d{\cal N}/dz$ via equation (4), can be obtained 
indirectly from the local galaxy luminosity function. 
We here adopt the $g$-band SDSS luminosity function from Montero-Dorta \& Prada (2009), 
who give Schechter parameters of $\alpha=-1.10$ and 
$\phi^{\star}=1.25\times10^{-2}\,h^3$\,Mpc$^{-3}$ for $h=1.0$.
We transform these parameters to the cosmological frame defined in Sect.\,3.3.
and calculate $\phi=\phi^{\star}\,\Gamma(\alpha+1,L/L^{\star})$ for different luminosity 
bins, where $\Gamma$ is the incomplete gamma function and $L^{\star}$
is the characteristic luminosity that characterizes the cut-off for the 
power-law component in the Schechter luminosity function (Schechter 1976).

%%%%%%%%%%%%%%%%%%%%% TABLE 1 %%%%%%%%%%%%%%%%%%%%%%

\begin{table*}[th!]
\caption[]{Models$^{\rm a}$ for absorber cross sections based on equation (2)}
\begin{tabular}{lrrrrrrr}
\hline
Model &  $\alpha$ & $\phi^{\star}_{\rm min}$ & $\phi^{\star}_{\rm pref}$ & $\phi^{\star}_{\rm max}$ &
$d{\cal N}/dz_{\rm min}$ & $d{\cal N}/dz_{\rm pref}$ & $d{\cal N}/dz_{\rm max}$ \\
          &           & [$h^3\,$Mpc$^{-3}$]      & [$h^3\,$Mpc$^{-3}$]       & [$h^3\,$Mpc$^{-3}$]      &
                         &                           &                          \\
\hline
1 & -1.00 & 0.0075 & 0.0125 & 0.0175 & 1.81 & 3.02 & 4.23 \\
2 & {\bf -1.10} & 0.0075 & {\bf 0.0125} & 0.0175 & 2.18 & {\bf 3.64} & 5.09 \\
3 & -1.20 & 0.0075 & 0.0125 & 0.0175 & 2.73 & 4.56 & 6.38 \\
4 & -1.30 & 0.0075 & 0.0125 & 0.0175 & 3.58 & 5.97 & 8.35 \\
\hline
\end{tabular}
\noindent
\\
{\small $^{\rm a}$\,The following Schechter parameters are given: $\alpha=$\,slope of the power-law component; 
$\phi^{\star}=\,$normalization density;\\ the preferred parameter combination is indicated with the
bold-face font.}
\end{table*}

%%%%%%%%%%%%%%%%%%%%%% FIGURE 08 %%%%%%%%%%%%%%%%%%%%%%

\begin{figure}[t!]
\resizebox{0.95\hsize}{!}{\includegraphics{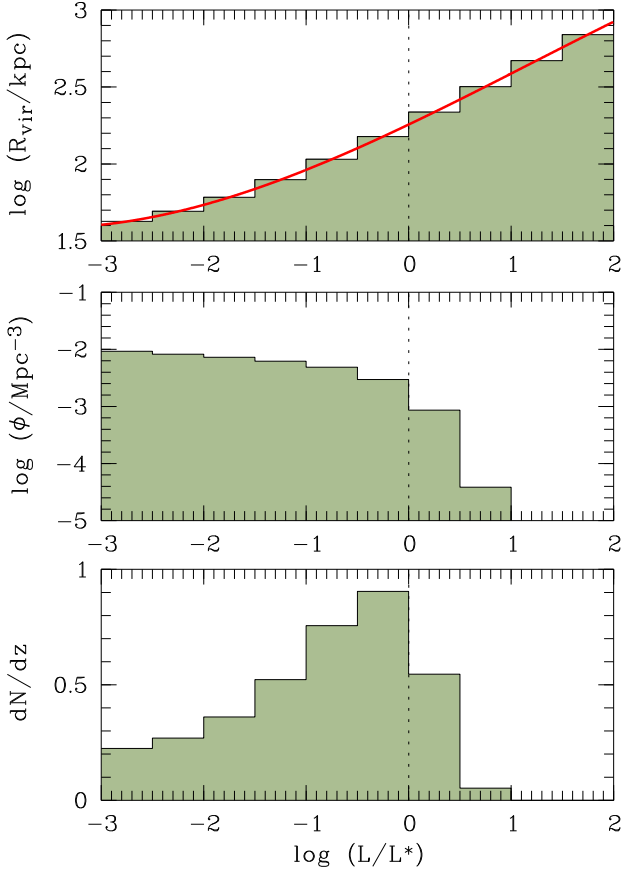}}
\caption[Halo model 1]{
Predictions from the halo model described in Sect.\,5.
{\it Upper panel:} The virial radius of galaxies as a function of
galaxy luminosity (adapted from Stocke et al.\,2014).
{\it Middle panel:} Galaxy space density as a function of
galaxy luminosity (from Montero-Dorta \& Prada 2009).
{\it Lower panel:} Expected number density of CGM absorbers as
a function of galaxy luminosity.
}
\end{figure}

%%%%%%%%%%%%%%%%%%%%%% FIGURE 09 %%%%%%%%%%%%%%%%%%%%%%

\begin{figure}[hb!]
\resizebox{0.95\hsize}{!}{\includegraphics{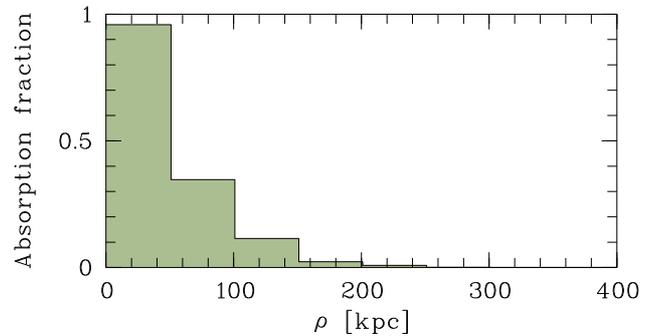}}
\caption[Halo model 2]{
CGM absorption fraction 
as a function of impact parameter, as predicted
from the halo model described in Sect.\,5.
}
\end{figure}

%%%%%%%%%%%%%%%%%%%%%%%%%%%%%%%%%%%%%%%%%%%%%%%%%%%%%%%%

The key assumption in our toy model is that all halos of galaxies with 
luminosities $L\geq 0.001\,L^{\star}$ contain metal-absorbing gas 
within their virial radii ($R_{\rm vir}$). We initially assume a 
unity covering fraction of the absorbing gas, so that 
the effective absorption cross section is simply $\pi\,R_{\rm vir}^2$.
Also the virial radius of a galaxy can be inferred indirectly from its
luminosity. As recently discussed by Stocke et al.\,(2014), 
halo matching models provide a relatively well-defined relation 
between $R_{\rm vir}$ and $L$ that can be used to define 
a scaling relation between these two parameters.
We define $C={\rm log}\,(L/L^{\star})$ and approximate 
the relation between $L$ and $R_{\rm vir}$ shown in Stocke et al.\,
(their Fig.\,8) by the fit

%%%%%%%%%%%%%%%%%%%%%%%%%%%%%%%%%%%%%%%%%%%%%%%%%%%%%%%%

\begin{equation}
{\rm log}\,R_{\rm vir}=
2.257+0.318\,C+
0.018\,C^2-
0.005\,C^3.
\end{equation}

%%%%%%%%%%%%%%%%%%%%%%%%%%%%%%%%%%%%%%%%%%%%%%%%%%%%%%%%

Equipped with these relations we can now assign to each luminosity bin a 
galaxy space density $\phi(L)$ and an effective absorption cross section
$A(L)=\pi\,R_{\rm vir}^2(L)$, which is all we need to calculate 
the expected number density of absorbers $d{\cal N}/dz(L)$ for the same bin. 
Integration over all galaxy luminosity bins then provides the expected 
total number density of absorbers arising from the halos of these galaxies.

Fig.\,8 summarizes the main results from our modeling. The red solid 
line in the upper panel visualizes the $L-R_{\rm vir}$-relation 
defined by equation (3), while the green-shaded histogram 
indicates the same relation in bins of log $(L/L^{\star})=0.5$. In the middle 
panel of Fig.\,8 the logarithmic galaxy density (log $\phi(L))$ is plotted
against log $(L/L^{\star})$ (same binning), while in the lower panel
we show the expected number density of absorbers $d{\cal N}/dz(L)$ 
(assuming a unity covering fraction, $\langle f_c \rangle=1$) 
as a function of log $(L/L^{\star})$ for the same binning. 
By integrating over the desired luminosity range we obtain an estimate
for the total number density of absorbers that arise in the gaseous halos
of the galaxies in that luminosity range.

While our observational galaxy sample is restricted to relatively bright
galaxies (see Sect.\,3), we here consider the much
larger range $L/L^{\star}\geq 0.001$ to evaluate
the potential contribution of faint galaxies and their halos to the total 
cross section of intervening metal absorbers, even if they remain
unseen in galaxy surveys. For $L/L^{\star}\geq 0.001$ and unity 
covering fraction we obtain a total absorber number density of $d{\cal N}/dz=3.6$. 
If we include the mean covering fraction as free parameter 
(which then represents an area-weighted mean), we can write more generally: 

%%%%%%%%%%%%%%%%%%%%%%%%%%%%%%%%%%%%%%%%%%%%%%%%%%%%%%%%

\begin{equation}
d{\cal N}/dz=3.6\,\langle f_{\rm c}\rangle.
\end{equation}

%%%%%%%%%%%%%%%%%%%%%%%%%%%%%%%%%%%%%%%%%%%%%%%%%%%%%%%%

For realistic gas covering fractions in galaxy halos in the range
$\langle f_{\rm c}\rangle=0.1-1.0$
this absorber number density is remarkably close to the observed 
number density of intervening Si\,{\sc iii} absorbers, as discussed
in the previous section.

%%%%%%%%%%%%%%%%%%%%% TABLE 2 %%%%%%%%%%%%%%%%%%%%%%

\begin{table}[h!]
\caption[]{Model predictions for effective CGM covering fractions, $\langle {\cal F}_c(r_{\rm max})\rangle$,
for different halo radii and galaxy luminosities (see Sect.\,5).}
\begin{footnotesize}
\begin{tabular}{rcccccc}
\hline
&   &  & \multicolumn{4}{c}{$\langle {\cal F}_c(r_{\rm max})\rangle$} \\
\hline
$r_{\rm max}$  & $\langle f_{\rm c}\rangle$ & & $L/L^{\star}$ & $L/L^{\star}$ & $L/L^{\star}$ & $L/L^{\star}$ \\
$[$kpc$]$      &                            & & $\geq 0.001$  & $\geq 0.01$   & $\geq 0.1$    & $\geq 1$      \\
\hline
200            & 1.00                       & & 0.18                    & 0.27                   & 0.48                  & 1.00           \\
400            & 1.00                       & & 0.04                    & 0.07                   & 0.12                  & 0.31           \\
1000           & 1.00                       & & 0.01                    & 0.01                   & 0.02                  & 0.05           \\
\hline
200            & 0.75                       & & 0.13                    & 0.20                   & 0.36                  & 0.96           \\
400            & 0.75                       & & 0.03                    & 0.05                   & 0.09                  & 0.23           \\
1000           & 0.75                       & & 0.01                    & 0.01                   & 0.01                  & 0.04           \\
\hline
200            & 0.50                       & & 0.09                    & 0.14                   & 0.24                  & 0.62           \\
400            & 0.50                       & & 0.02                    & 0.03                   & 0.06                  & 0.16           \\
1000           & 0.50                       & & 0.00                    & 0.01                   & 0.01                  & 0.02           \\
\hline
\end{tabular}
\noindent
\\
\end{footnotesize}
\end{table}

%%%%%%%%%%%%%%%%%%%%%%%%%%%%%%%%%%%%%%%%%%%%%%%%%%%%%%%%

One serious concern about the relevance of the above given estimate is, how strongly 
the derived absorber number density depends on the input parameters from
the adopted galaxy luminosity function. To explore this dependence we have
calculated $d{\cal N}/dz$ for a whole set of (realistic) Schechter parameters,
as presented in Table 1. For this, we vary for a given slope $\alpha$ in each row
the normalization density from $\phi^{\star}_{\rm min}$ over $\phi^{\star}_{\rm pref}$ to
$\phi^{\star}_{\rm max}$ (where $\phi^{\star}_{\rm pref}=0.0125\,h^3$ Mpc$^{-3}$ 
represents our preferred normalization; see above) and calculate the expected 
absorber number densities $d{\cal N}/dz_{\rm min}$, $d{\cal N}/dz_{\rm pref}$,
$d{\cal N}/dz_{\rm max}$. For Schechter parameters in the range $\alpha=[-1.0,-1.3]$
and $\phi^{\star}=[0.0075,0.150]$ the expected number densities vary between 
$d{\cal N}/dz=1.81$ and $8.35$. It is therefore valid to claim that
the expected contribution of the metal-enriched CGM to the cross section
of intervening metal absorbers is $d{\cal N}/dz<9$ (this 
conclusion holds for any metal ion that can be observed via QSO absorption
spectroscopy).

One may argue that the values for $d{\cal N}/dz$ estimated in this 
manner over-estimate the true absorption-cross section of gaseous halos
because galaxy halos can overlap with each other (in particular 
satellite galaxies and their hosts), so that the effective halos cross 
section of a galaxy ensemble would be smaller than the sum of $A$ for all 
individual galaxies in that ensemble. This effect is commonly investigated 
in studies that aim at characterizing the relationship between the stellar
masses of galaxies and their superordinate dark matter (DM) halos, e.g., by using
statistical methods such as the Halo Occupation Distribution (HOD) formalism
and others (e.g., Peacock \& Smith 2000; Moster et al.\,2010). From these
models follows that, on average, only the most massive halos (with DM halo masses
$>10^{11.8}\,M_{\sun}$, corresponding to $L>2L^{\star}$) are occupied by more 
than one galaxy. These halos are rare, however, and their contribution 
to $d{\cal N}/dz$ is small (see Fig.\,8), so that a small change in the 
absorption cross section of such massive halos basically would have no 
influence on the estimate of the integrated number density of intervening
absorbers.

A comparison between the estimated gas cross section of galaxy halos from
our model with the observed number density of intervening Si\,{\sc iii} absorbers
allows us to draw some very interesting conclusions.
The first important conclusion is that, {\it because 
$d{\cal N}/dz($Si\,{\sc iii}$)<d{\cal N}/dz($model) 
for log $N$(Si\,{\sc iii}$)>12.2$, the entire population of these Si\,{\sc iii} 
systems can be explained by metal-enriched gas that is gravitationally bound to
galaxies}, i.e., there is no need for an intergalactic gas component that
hosts Si\,{\sc iii} absorbers at this strength. Turning this argument
around, our model also predicts that intervening metal absorbers with 
number densities $d{\cal N}/dz>9$ (e.g., O\,{\sc vi} absorbers with
$d{\cal N}/dz\approx 16$; Tripp et al.\,2008) must partially arise from
gas outside of the virial radius of galaxies. This aspect is further
discussed in Sect.\,8.
A second conclusion is that for our preferred model with 
$\alpha=-1.1$ and $\phi^{\star}=0.0125\,h^3$ Mpc$^{-3}$ 
(Montero-Dorta \& Prada 2009) the mean covering fraction of 
Si\,{\sc iii} within $R_{\rm vir}$ is expected to be 
$\langle f_c \rangle=2.5\pm 0.4/3.6=0.69\pm 0.11$, a value that is 
similar to the covering fraction of Si\,{\sc iii} in the halo of 
the Milky Way, as recently derived from QSO absorption-line observations
(Herenz et al.\,2013; Lehner et al.\,2012). In Sect.\,6
we further investigate, whether the {\it observed} distribution of galaxies around
Si\,{\sc iii} absorbers is in line with these conclusions.

Our model further allows us to predict the mean absorption fraction around
galaxies as a function of impact parameter $\rho$. 
We here define the mean absorption fraction as
the detection rate of intervening absorption from CGM gas for a given
impact parameter for a population of galaxies with different virial
radii. Only the most luminous galaxies with large $R_{\rm vir}$ are expected
to contribute to the mean absorption at $\rho>100$ kpc, while for small
impact parameters $\rho<50$ kpc both luminous as well as faint galaxies do contribute.
Consequently, the mean absorption fraction around galaxies is expected to 
decline in a way that is determined by the distribution of $R_{\rm vir}$
(and thus by the shape of the galaxy luminosity function).

In Fig.\,9 we show the mean absorption fraction plotted against 
$\rho$ (in bins of 50 kpc), as derived from our preferred model
with Schechter parameters $\alpha=-1.1$ and $\phi^{\star}=0.0125\,h^3$ Mpc$^{-3}$.
As expected, the absorption rate decreases rapidly with increasing impact 
parameter because the fraction of sightlines that pass galaxies at $\rho>R_{\rm vir}$
increases. Fig.\,9 can be directly compared to the observed 
absorption rate around low-redshift galaxies, as is presented 
in Sect.\,6. 

Another important piece of information that can be extracted from our model concerns
the expected covering fraction of circumgalactic gas around a population
of galaxies in a given luminosity range.
It is useful to transform the absorption rate into an {\it effective covering fraction},
$\langle {\cal F}_c(r_{\rm max})\rangle$, that is normalized to a {\it fixed} radius 
$r_{\rm max}$ (instead of being normalized to $R_{\rm vir}$, which depends 
on the galaxy's mass and luminosity). To carry out this transformation we
need to take into account that the covering fraction describes the detection rate
per unit {\it area}. Since the area of each ring with thickness
$\Delta r$ and radius $r$ is $\Delta r\,2\pi r$, the
absorption rates need to be weighted with $r$ and integrated
from $r=0$ to $r=r_{\rm max}$ to obtain $\langle {\cal F}_c(r_{\rm max})\rangle$.
In this way, it is possible to predict from our model the effective covering fractions
for different values of $r_{\rm max}$ for a population of galaxies in a given 
luminosity range.
In Table 2 we list the predicted effective CGM covering fractions for different 
values of $r_{\rm max}$, $\langle f_c\rangle$, and different luminosity ranges based on the 
preferred parameters for the galaxy luminosity function (see Table 1).
We would like to emphasize again that in this approach $\langle {\cal F}_c \rangle$ describes the average 
CGM covering fraction of a population of galaxies within an annulus with a fixed 
radius, while $\langle f_c\rangle$ describes the mean covering fraction of the galaxies' 
CGM within $R_{\rm vir}$.

%%%%%%%%%%%%%%%%%%%%%% FIGURE 10 %%%%%%%%%%%%%%%%%%%%%%

\begin{figure*}[t!]
\begin{center}
\resizebox{1.0\hsize}{!}{\includegraphics{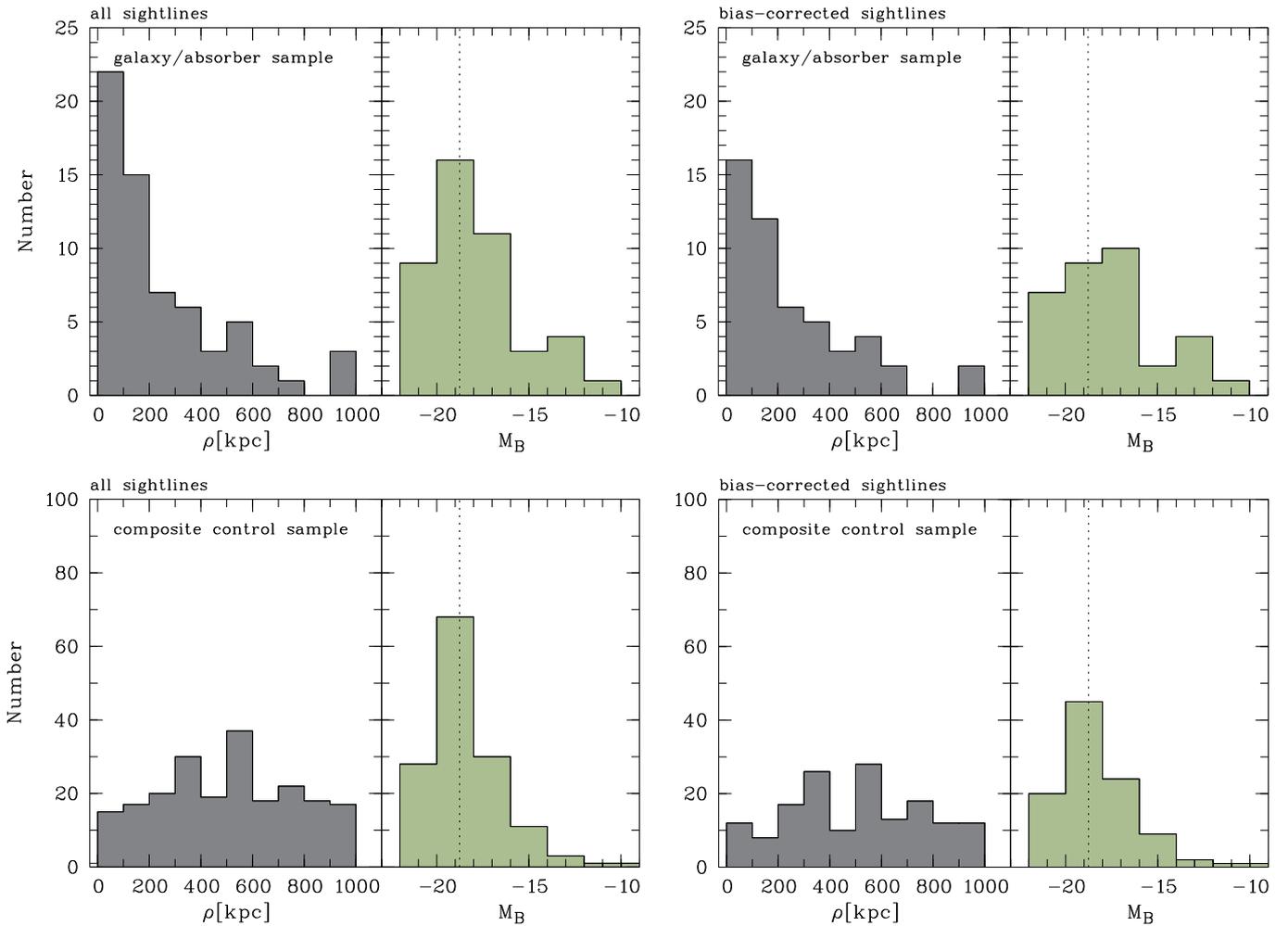}}
\caption[Absorber-galaxy relation]{
{\it Upper panels:} Distributions of impact parameters and
absolute $B$-band magnitudes for galaxies that are
located within $|\Delta v|=1000$ km\,s$^{-1}$ of the
absorbers (galaxy/absorber samples). On the left-hand side
we show the distribtions for all sightlines that exhibit 
Si\,{\sc iii} absorption, while on the right-hand side
only the bias-corrected set of sightlines is considered
(see Sect.\,4.2). Galaxies tend to cluster around the
LOS in velocity regions where intervening absorption is observed.
{\it Lower panels:} The same distributions for galaxies that
have no absorption systems within $|\Delta v|=1000$ km\,s$^{-1}$
(composite control samples; see Sect.\,6.2 for details). No 
clustering is observed in this case.
}
\end{center}
\end{figure*}

%%%%%%%%%%%%%%%%%%%%%%%%%%%%%%%%%%%%%%%%%%%%%%%%%%%%%%%%

\section{Observed absorber-galaxy connection}

\subsection{Individual galaxies associated with Si\,{\sc iii} absorbers}

To characterize the true absorber-galaxy connection at $z\leq 0.1$ we 
have calculated the geometrical impact parameters $\rho$ between the 
sightlines that exhibit Si\,{\sc iii} absorbers at $z\leq 0.1$ in our COS data
and nearby galaxies in our galaxy catalog (see Sect.\,3.3).

Ideally, one would study the relation between absorbers and galaxies 
using the absorber-galaxy cross-correlation function together with a 
well-defined galaxy completeness function. Our galaxy sample, that is 
based on many different galaxy surveys and individual observations, 
is highly inhomogeneous with respect to the completeness limit at the 
faint end of the $B$ band magnitude distribution, however, and therefore 
a statistically meaningful galaxy completeness function cannot be obtained.
We would like to point out, however, that the motivation for including these galaxy data in our study 
is not to provide a statistically complete sample of absorber/galaxy pairs, but 
to qualitatively check the plausibility of our hypothesis, in which most 
Si\,{\sc iii} absorbers at $z\leq 0.1$ arise in the extended
gaseous halos of luminous low-redshift galaxies, for which the present galaxy 
data sample is sufficient in depth.

Because of the above mentioned limitations 
we cannot simply combine the impact-parameter 
distributions of the individual sightlines to statistically investigate 
the overall absorber-galaxy connection, but instead have to evaluate the 
connection between absorbers and galaxies along each sightline 
{\it individually} before combining the results.
In the following, we further specify our strategy how to do this.

For each sightline that contains an intervening Si\,{\sc iii} absorber 
we first selected all galaxies that have recession velocities within 
$|\Delta v|=1000$ km\,s$^{-1}$ of the absorber (hereafter referred 
to as "LOS absorber sample") and identified the galaxy with the 
smallest impact parameter to the sightline. For the same line of 
sight we then created a set of nine galaxy control samples
covering the adjacent velocity range $|\Delta v|=1000-10,000$ km\,s$^{-1}$,
where each LOS control sample spans a velocity range of 
$|\Delta v|=1000$ km\,s$^{-1}$, similar as for the LOS absorber sample. 
We again marked the galaxies that are nearest to the LOS.
The galaxy samples constructed in this way have the same magnitude 
completeness for each sightline.

If intervening absorbers would trace the CGM of galaxies that are close 
to the sightlines (as we suspect), one would expect for a large-enough 
absorber/galaxy sample (as provided here) that the impact parameters of 
the nearest galaxy to the LOS in the absorber samples are, on average, 
substantially smaller than in the control samples. Such a trend is indeed 
seen in our data: for 76 percent of the LOS that exhibit an intervening 
Si\,{\sc iii} absorber, the galaxy with the smallest impact parameter 
is located within $1000$ km\,s$^{-1}$ of the absorption redshift.

In the upper panels of Fig.\,10 (gray-shaded histograms) we show the distribution 
of $\rho$ for the nearest galaxies in the LOS absorber samples ("galaxy/absorber sample")
for all sightlines with Si\,{\sc iii} absorbers (left column) and for
the bias-corrected set of sightlines with Si\,{\sc iii} absorbers (right column).
Obviously, galaxies cluster around sightlines if an Si\,{\sc iii} absorber is
present. This trend is evident in both galaxy/absorber samples (total sample
and bias-corrected sample), but in the (biased) total sample there are more
data points in the lowest $\rho$-bin, as is expected from the design of the
Bordoloi et al.\,(2014) COS-Dwarfs survey (see Sect.\,4.2).

Still, in the bias-corrected galaxy/absorber sample, $56$ percent of the galaxies 
are located at impact parameters $\rho<200$ kpc and for the same percentage the velocity 
difference between the galaxy and the absorber is $|\Delta v|\leq 300$ km\,s$^{-1}$.
From our model we would expect that the maximum impact parameter in an idealized
galaxy sample is $\rho\approx300$ kpc, as this length scale corresponds to the virial
radius of the most luminous galaxies that are expected to significantly contribute to
the number density of Si\,{\sc iii} absorbers before the galaxy-density distribution
breaks down for higher luminosities (Fig.\,8, middle panel).
However, since our galaxy sample is incomplete and highly inhomogeneous along the 
different LOS with respect to galaxies at the faint end of the luminosity function,
the observed distribution of $\rho$ extends much further out.
With the green-shaded histogram we show in Fig.\,10 the absolute $B$-band magnitudes for the 
galaxies in the galaxy/absorber samples, as calculated from their
cosmological distance. The dotted line indicates the absolute $B$ magnitude of an
$L^{\star}$ galaxy (adopted from Montero-Dorta \& Prada 2009).

%%%%%%%%%%%%%%%%%%%%%% FIGURE 11 %%%%%%%%%%%%%%%%%%%%%%

\begin{figure}[t!]
\begin{center}
\resizebox{0.9\hsize}{!}{\includegraphics{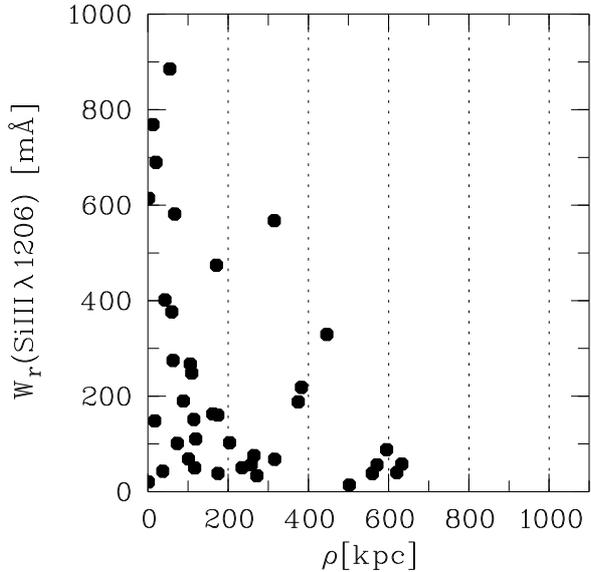}}
\caption[EW galaxies]{Restframe equivalent widths of Si\,{\sc iii}
$\lambda 1206$ are plotted against the impact parameters of the
galaxies nearest to the absorbers. For increasing impact parameters the maximum
value for $W_{\rm r}$ is decreasing from $W_{\rm r}\approx 900$ m\AA\,down to values
$<100$ m\AA.
}
\end{center}
\end{figure}

%%%%%%%%%%%%%%%%%%%%%% FIGURE 12 %%%%%%%%%%%%%%%%%%%%%%

\begin{figure}[t!]
\resizebox{1.0\hsize}{!}{\includegraphics{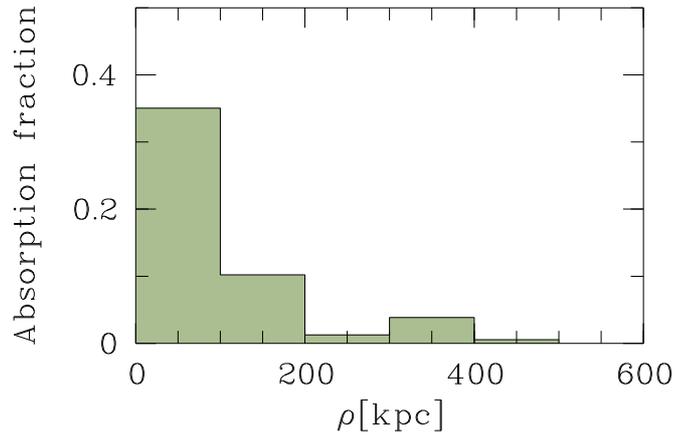}}
\caption[Absorption fraction]{
Absorption fraction (fractional abundance of intervening absorbers
per galaxy) as a function of impact parameter, as measured
in our galaxy/absorber sample.
}
\end{figure}

%%%%%%%%%%%%%%%%%%%%%%%%%%%%%%%%%%%%%%%%%%%%%%%%%%%%%%%%

To check the statistical significance of this clustering trend we 
analyzed the galaxies in the respective control samples in a 
similar manner. We find that the galaxies do not cluster
around the sightlines in any of the control samples, but their impact 
parameters are more or less evenly distributed. 
To demonstrate this, we show in the lower left and right
panels of Fig.\,10  the impact-parameter distributions in the control
samples, i.e., for galaxies that have no absorption systems within 
$|\Delta v|=1000$ km\,s$^{-1}$. For this, we combined for 
each of the two data sets (total sample/bias-corrected sample) the 
nine individual control samples to one "composite control sample", 
respectively, which contains all galaxies within the velocity window 
$|\Delta v|=1000-10,000$ km\,s$^{-1}$ with respect to the absorber
redshift (see above). 

The distributions of absolute $B$-band magnitudes in the 
composite control samples (Fig.\,10, lower left panel, green-shaded areas) 
are similar to those in the galaxy/absorber samples, proving that the
different galaxy samples have the same completeness in $B$. 

In summary, the observed absorber-galaxy relation strongly suggests that 
the galaxies "know" about the presence of nearby Si\,{\sc iii} 
absorption systems. This is exactly what would be expected if the 
absorption were (predominantly) caused by metal-enriched gas 
in the extended halos and the superordinate cosmological environment of 
these galaxies. 

For our bias-corrected galaxy/absorber sample we calculate effective covering 
fractions of $\langle {\cal F}_c(r_{\rm max})\rangle=0.08$ for $r_{\rm max}=400$ 
kpc and $\langle {\cal F}_c(r_{\rm max})\rangle=0.01$ for $r_{\rm max}=1000$ kpc.
These effective covering fractions are comparable to those expected for 
a population of $L>0.1L^{\star}$ galaxies that are surrounded by a CGM 
that reaches out to $R_{\rm vir}$ with $\langle f_c \rangle=0.75$
(Table 2).

\subsection{Absorption strength vs.\,impact parameter}

In Fig.\,11 we plot the restframe equivalent widths of Si\,{\sc iii}
$\lambda 1206$ versus the impact parameters of the galaxies nearest to
the absorbers. For $\rho<200$ kpc the equivalent width scatters strongly
in the range $10-1000$ m\AA. For increasing impact parameters the maximum
value for $W_{\rm r}$ (hereafter referred to as $W_{\rm r,max}$) is decreasing 
to values $<100$ m\AA\, for $\rho>600$ kpc. 

The interpretation of this plot is not as simple as it may look like: 
because of the incompleteness of the galaxy sample, the largest values 
for $\rho$ (which are far beyond the virial radii of
the most massive galaxies) do not represent true impact parameters
to the nearest galaxies, but rather indicate the impact parameters to
the nearest {\it luminous} galaxies. From observations of strong 
Mg\,{\sc ii} absorbers it is known that the mean 
Mg\,{\sc ii} $\lambda 2796$ equivalent width in the CGM of luminous
galaxies is larger than for low-luminosity galaxies (e.g., Nielsen
et al.\,2013). Assuming that a similar trend holds for Si\,{\sc iii}, 
the large scatter for $W_{\rm r}$(Si\,{\sc iii}\,$\lambda 1206$) at
$\rho<200$ kpc can be interpreted by the large range in luminosities 
of the galaxies whose circumgalactic gas causes the absorption, where 
$W_{\rm r,max}$ is is determined by the most luminous galaxies.
The decline of $W_{\rm r,max}$ for increasing $\rho$ then simply 
reflects the decreasing likelihood to {\it miss} luminous galaxies 
around the absorbers that are responsible for for the strongest
absorbers. For large values of $\rho$, only absorbers with small
$W_{\rm r}$ values remain, as they belong to low-luminosity galaxies
that are too faint to be covered in our galaxy sample.

\subsection{QSO sightlines associated with individual galaxies}

We now investigate the absorber/galaxy-connection from the perspective of the
galaxies. For each galaxy in our sample we first identified QSO sightlines that pass
the galaxies at impact parameters $\rho<1000$ kpc. 
We then further selected only galaxies in redshift ranges that are covered by each
relevant COS spectrum at a S/N that is high enough to detect Si\,{\sc iii} absorption at 
column densities log $N_{\rm lim}$(Si\,{\sc iii}$)\geq12.2$. Finally, we calculated for
each $100$ kpc wide impact-parameter bin the absorption fraction (i.e., the fraction of 
galaxies having a Si\,{\sc iii} absorber above this column density limit in
this impact-parameter range). The result from this analysis is shown in 
Fig.\,12.
As can be seen, the observed absorption fraction is 
substantially smaller than the one derived from the 
idealized halo model (Fig.\,9), but extends to much larger impact parameters even 
beyond the expected virial radii of massive galaxies.
We again interpret this behavior as a sign for the incompleteness in our 
galaxy data, i.e., we suspect that sightlines that have apparent galaxy 
impact parameters in the range $300\leq \rho <1000$ kpc are arising from halo
gas within the virial radii of {\it unseen} galaxies.

%%%%%%%%%%%%%%%%%%%%%%%%%%%%%%%%%%%%%%%%%%%%%%%%%%%%%%%%

\section{Ionization conditions}

\subsection{Model setup}

To gain insight into the physical properties of
the intervening Si\,{\sc iii} absorbers and to estimate the 
total gas mass that they trace we studied
the ionization conditions in the absorbers using the
ionization code Cloudy (v13.03; Ferland et al.\,2013).
The absorbers are modelled as plane-parallel
slabs with fixed neutral gas column densities;
they are exposed to the UV background radiation with 
log $\Gamma=-13.6$ (see Sect.\,2 for more details on
the used UV background field) and are assumed to be 
optically thin in H\,{\sc i}.

%%%%%%%%%%%%%%%%%%%%%% FIGURE 13 %%%%%%%%%%%%%%%%%%%%%%

\begin{figure}[t!]
\resizebox{1.0\hsize}{!}{\includegraphics{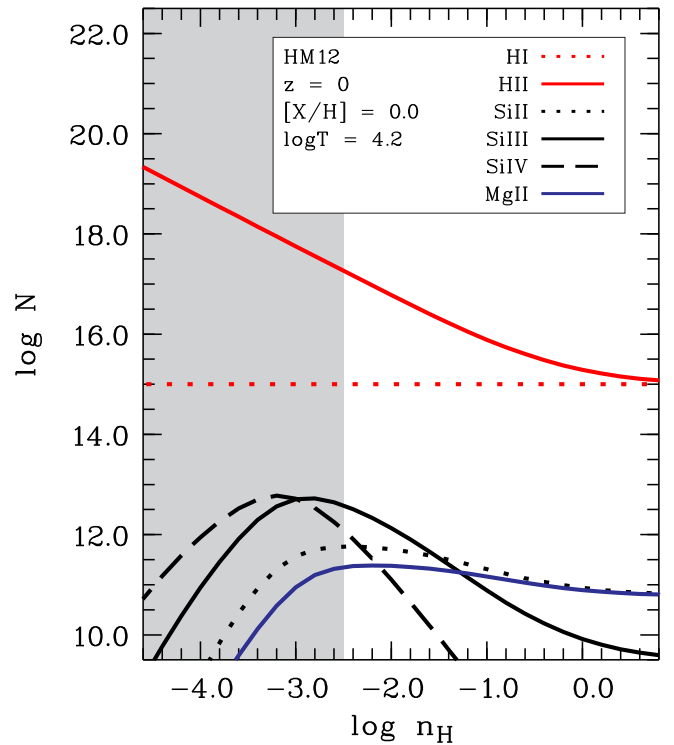}}
\caption[Cloudy modeling]{
Predictions of H\,{\sc ii}, Si\,{\sc ii}, Si\,{\sc iii}, Si\,{\sc iv}, and
Mg\,{\sc ii} column densities as
a function of the gas density (log $n_{\rm H}$) for an
H\,{\sc i} column density of log $N$(H\,{\sc i}$)=15$,
based on a Cloudy ionization model using the local UV background 
and assuming absolute and relative solar
abundances of all elements. The gray-shaded area indicates the density
that is typical for the Si\,{\sc iii}/Si\,{\sc iv} phase (see Sect.\,7.3).
}
\end{figure}

%%%%%%%%%%%%%%%%%%%%%%%%%%%%%%%%%%%%%%%%%%%%%%%%%%%%%%%%

In Sect.\,4 we have suggested that Si\,{\sc iii} traces 
at least two characteristic gas phases, a more ionized, diffuse
phase that is also traced by Si\,{\sc iv}, and a more dense
(partly neutral) phase that is also detected in Si\,{\sc ii}.
For our Cloudy modeling we therefore have focused on the relation between
the column densities of H\,{\sc i}, Si\,{\sc ii}, Si\,{\sc iii}, and
Si\,{\sc iv} as a function of the ionization parameter $U$,
the ratio between ionizing photon density
and total particle density (i.e., $U=n_{\gamma}/n_{\rm H}$).
For an assumed ionizing radiation field one can calculate
$n_{\gamma}$ and thus can relate U with the gas density $n_{\rm H}$.

In Fig.\,13 we have plotted (as an example) the expected 
H\,{\sc ii}, Si\,{\sc ii}, Si\,{\sc iii}, and Si\,{\sc iv}
column densities against log $n_{\rm H}$ for a (typical)
H\,{\sc i} column density of log $N$(H\,{\sc i}$)=15$,
based on a set of Cloudy models assuming solar abundances of all heavy 
elements (Asplund et al.\,2009).
We also show the expected trend for Mg\,{\sc ii}, which follows Si\,{\sc ii}
very closely over the entire density range. This similarity justifies our
previous approach to estimate the number density of strong Mg\,{\sc ii} using
Si\,{\sc ii} as a proxy (Sect.\,4.3.2). 
We have set up a grid of Cloudy models in which we have varied the H\,{\sc i}
column density in the range log $N$(H\,{\sc i}$)=14-19$ to provide 
column-density predictions for the above listed ions.

The most important conclusion from this modeling is
that the shapes of the column-density curves for H\,{\sc ii}, Si\,{\sc ii}, Si\,{\sc iii},
Si\,{\sc iv}, and Mg\,{\sc ii} (and their positions relative to each other) basically 
do not depend on log $N$(H\,{\sc i}), implying that the observed column density 
ratios of the Si ions can be used to constrain $n_{\rm H}$ in optically thin
H\,{\sc i}/Si\,{\sc iii} absorbers even without knowing $N$(H\,{\sc i}).

Fig.\,13 further indicates that Si\,{\sc iv} becomes the dominant ion of
Si only for relatively low gas densities (log\,$n_{\rm H}<-3.5$), while 
Si\,{\sc iii} dominates in the density range $-3.5<$\,log\,$n_{\rm H}\leq -2.0$.
Si\,{\sc ii} is dominant at log\,$n_{\rm H}>-2.0$, thus at densities that are
expected to be relevant only for DLAs and sub-DLAs in the neutral gas disks
and in predominantly neutral gas structures in the inner halos of galaxies
(e.g, in high-velocity clouds).

\subsection{Single-phase model}

One may assume that the simultaneous absorption of Si\,{\sc ii}, 
Si\,{\sc iii}, and Si\,{\sc iv} at similar radial velocities, 
such as observed in some systems, stems from a 
single gas phase in the absorbers. For that case, our Cloudy models provide
some firm predictions for the allowed column-density ranges for these
three ions that can be summarized by the following parametrization of 
the expected column-density ratios Si\,{\sc iii}/Si\,{\sc iv} and 
Si\,{\sc iii}/Si\,{\sc ii} in a single gas phase:

%%%%%%%%%%%%%%%%%%%%%%%%%%%%%%%%%%%%%%%%%%%%%%%%%%%%%%%%

\begin{equation}
{\rm log}\,\left[\frac{N({\rm Si\,III})}{N({\rm Si\,IV})}\right]=1.68-
1.44\,{\rm log}\,\left[\frac{N({\rm Si\,III})}{N({\rm Si\,II})}\right].
\end{equation}

%%%%%%%%%%%%%%%%%%%%%%%%%%%%%%%%%%%%%%%%%%%%%%%%%%%%%%%%

In Fig.\,14 we have plotted the observed column-density ratios of
these ions (and relevant limits) together with predictions from the
Cloudy model as given in equation (5) (the model is indicated 
in Fig.\,14 by the black solid line). For none of the systems that have measured
column densities for Si\,{\sc ii}, Si\,{\sc iii}, and Si\,{\sc iv}
(Fig.\,14 filled circles) do the data points lie on the expected
relation for the single-phase model. For all systems for which we have 
measured values in these three ions the 
Si\,{\sc ii} and/or Si\,{\sc iv} column densities are too high for the
observed Si\,{\sc iii} column density to match the single-phase model. 
While we cannot exclude that at least some of the sytems for which 
only lower limits for Si\,{\sc iii}/Si\,{\sc iv} and Si\,{\sc iii}/Si\,{\sc ii} 
are available are in accordance with
the single-phase model, the observations clearly do not favor 
such a scenario, but rather point towards a more complex multiphase
nature of gas, as considered below.

%%%%%%%%%%%%%%%%%%%%%% FIGURE 14 %%%%%%%%%%%%%%%%%%%%%%

\begin{figure}[t!]
\begin{center}
\resizebox{0.9\hsize}{!}{\includegraphics{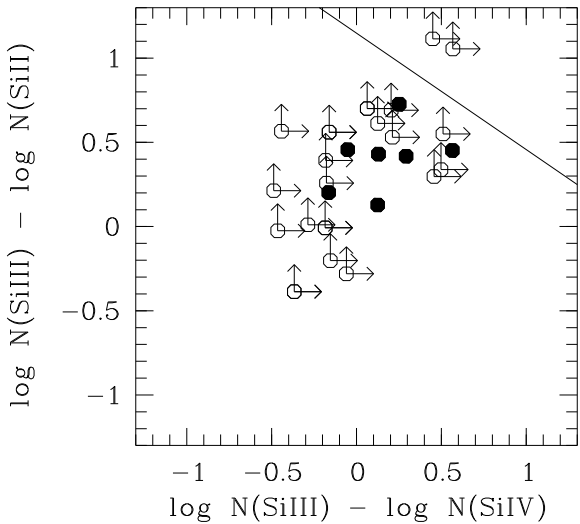}}
\caption[Gas phase ratios]{
Column-density ratios of Si\,{\sc iii}/Si\,{\sc ii} vs.\,
Si\,{\sc iii}/Si\,{\sc iv} are plotted together with predictions from
the single-phase Cloudy model (equation 5, indicated by the black solid line).
Measured values are plotted with filled circles, limits
are indicated with open circles. For none of the systems
do the data points lie on the expected relation for the single-phase model,
implying that the gas is multiphase.
}
\end{center}
\end{figure}

%%%%%%%%%%%%%%%%%%%%%% FIGURE 15 %%%%%%%%%%%%%%%%%%%%%%

\begin{figure}[t!]
\begin{center}
\resizebox{0.8\hsize}{!}{\includegraphics{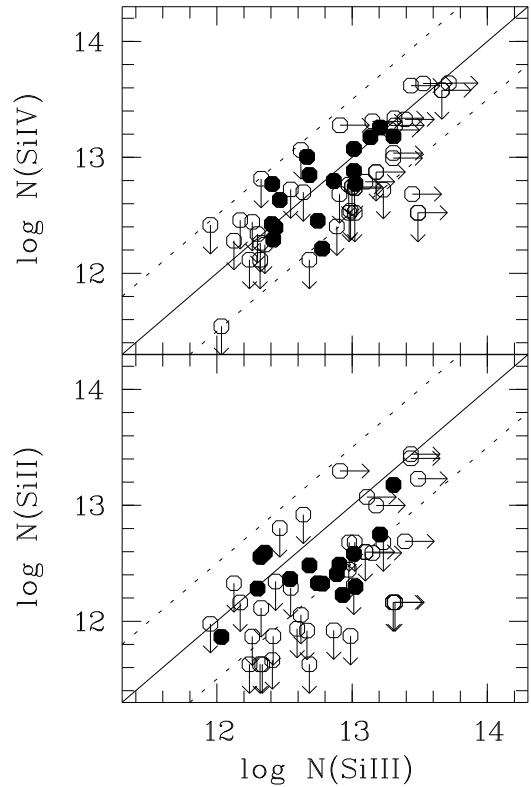}}
\caption[Silicon ratio]{
Distribution of logarithmic column densities in the Si\,{\sc iii} absorbers.
{\it Upper panel:} Si\,{\sc iv} vs.\,Si\,{\sc iii} (absolute values indicated
with filled circles, limits with open circles). {\it Lower panel:} Si\,{\sc ii} vs.\,Si\,{\sc iii}.
The dotted/solid lines indicate
ratios of $+0.5,0,-0.5$ dex, respectively.
}
\end{center}
\end{figure}

%%%%%%%%%%%%%%%%%%%%%%%%%%%%%%%%%%%%%%%%%%%%%%%%%%%%%%%%

\subsection{Multi-phase model}

The alternative (and probably more realistic) model for the absorbers is
that of a multiphase gas, where Si\,{\sc ii} and Si\,{\sc iv} predominantly
trace different gas phases (and different physical regions) that coexist within
the same overall absorbing gas structures. In the Milky Way halo, the 
existence of multiphase halo gas as traced be various low, intermediate,
and high ions is well established (e.g., Sembach et al.\,2003; 
Fox et al.\,2006; Collins et al.\,2009; Shull et al.\,2009; Richter
et al.\,2009; Herenz et al.\,2013). 

For the following we assume that Si\,{\sc iii} absorption arises in
both a diffuse ionized gas phase traced by Si\,{\sc iv}/Si\,{\sc iii}
as well as a somewhat denser (partly neutral) gas phase traced by 
Si\,{\sc ii}/Si\,{\sc iii}.
To further investigate the characteristic densities of these two
phases we have plotted in Fig.\,15 the column densities 
log $N$(Si\,{\sc iii}) vs.\, log $N$(Si\,{\sc iv}) (upper panel)
and log $N$(Si\,{\sc iii}) vs.\, log $N$(Si\,{\sc ii}) (lower panel).
Data points from measured column density limits in these ions
are plotted with open circles.
Because both gas phases defined above may be present in an absorber,
the plotted values of log $N$(Si\,{\sc iii}) have to be regarded
as an upper limit for the Si\,{\sc iii} column {\it in each phase}.

For the absorbers that are detected simultaneously in Si\,{\sc iii}
and Si\,{\sc iv} (upper panel of Fig.\,15, filled circles) 
the data points scatter within $\sim 0.7$ dex around the 
$N$(Si\,{\sc iii}$)=N$(Si\,{\sc iv}) line. 
For the absorbers that are detected simultaneously in Si\,{\sc iii}
and Si\,{\sc ii} (lower panel of Fig.\,15, filled circles) the
measured Si\,{\sc ii} column densities are typically lower than
that of Si\,{\sc iii}, but (again) note that a considerable fraction
of the Si\,{\sc iii} column may arise in the Si\,{\sc iv} phase,
so that log [$N$(Si\,{\sc ii})/$N$(Si\,{\sc iii})] could be much
higher locally. 

Because of the unknown intrinsic structure of each absorber it 
is challenging to provide firm predictions for the gas density 
for each individual system that has measured Si\,{\sc ii}/Si\,{\sc iii}/Si\,{\sc iv}
column densities. From the exploration of the parameter space the
Cloudy models deliver, however, a characteristic gas density 
that separates the Si\,{\sc iv}/Si\,{\sc iii} from the 
Si\,{\sc ii}/Si\,{\sc iii} phase in the absorbers (see also Fig.\,13).
We find that the Si\,{\sc iv}/Si\,{\sc iii} phase traces 
gas with densities log $n_{\rm H}\leq -3.0$ (in Fig.\,13 indicated with the 
gray-shaded area), while the Si\,{\sc ii}/Si\,{\sc iii} phase 
has higher densities in the range $n_{\rm H}>-3.0$.

\subsection{Total gas mass and baryon budget}

Our observations and Cloudy models imply that Si\,{\sc iii} traces 
diffuse (predominantly ionized) gas in the extended gaseous halos
(i.e., in the circumgalactic medium) of galaxies.
An interesting question is, how much mass is contained in such gas
and what is the overall baryon budget of intervening 
Si\,{\sc iii} absorbers and the metal-enriched CGM. 
To derive the total gas mass of the Si\,{\sc iii} systems in
our sample we need to calculate the amount of ionized hydrogen in
each absorber. Because of the much higher ionization fraction
in the Si\,{\sc iv}/Si\,{\sc iii} systems compared to the 
Si\,{\sc ii}/Si\,{\sc iii} phase we here concentrate on 
the estimate of $N$(H\,{\sc ii}) in low-density 
Si\,{\sc iii} absorbers that are associated with Si\,{\sc iv}.

For the range log $N$(H\,{\sc i}$)=14-17$ our Cloudy model
grids imply a relatively simple relation between the
minimum H\,{\sc ii} column density and the Si\,{\sc iii}
column density in each absorber:

%%%%%%%%%%%%%%%%%%%%%%%%%%%%%%%%%%%%%%%%%%%%%%%%%%%%%%%%

\begin{equation}
{\rm log}\,N({\rm H}\,{\rm II})\approx\,
{\rm log}\,N({\rm Si}\,{\rm III})
+{\rm log}\,Y-{\rm log}\,{\rm (Si/H)}_{\sun},
\end{equation}

%%%%%%%%%%%%%%%%%%%%%%%%%%%%%%%%%%%%%%%%%%%%%%%%%%%%%%%%

where the parameter $Y$, that solely depends on the gas density, needs 
to be determined from the observed Si\,{\sc iv}/Si\,{\sc iii}
ion ratios (Fig.\,13). Since we do not know what fraction of the Si\,{\sc iii}
column density can be assigned to the Si\,{\sc iv} phase, we can
only place a lower limit to $Y$ (see above).
Similarly, because of the unknown metallicity of the CGM, we have to assume
an upper limit for the silicon abundance (Si/H) in the gas. 
Equation (6) allows us to derive a lower limit for
$N$(H\,{\sc ii}) in each Si\,{\sc iv}/Si\,{\sc iii} absorber, from
which the integrated (=total) column density, $N$(H\,{\sc ii}$)_{\rm tot}$,
can then be determined.

The cosmological mass density of the Si\,{\sc iv}/Si\,{\sc iii} absorbers
in terms of the current critical density, $\rho_{\rm c}$, can
be estimated by

%%%%%%%%%%%%%%%%%%%%%%%%%%%%%%%%%%%%%%%%%%%%%%%%%%%%%%%%

\begin{equation}
\Omega_b{\rm (Si\,III)} \equiv \frac{\mu\,m_{\rm H}\,H_0}
{\rho_{\rm c}\,c}\,N({\rm H\,II})_{\rm tot}\,
\Delta X_{\rm tot}^{-1},
\end{equation}

%%%%%%%%%%%%%%%%%%%%%%%%%%%%%%%%%%%%%%%%%%%%%%%%%%%%%%%%

with $\mu=1.3$, $m_{\rm H}=1.673 \times 10^{-27}$ kg, $H_0=
69.7$ km\,s$^{-1}$\,Mpc$^{-1}$ (Hinshaw et al.\,2013), and $\rho_{\rm c}=3H_0\,^2/8 \pi G$.
The comoving path length $\Delta X$ available for the detection of
Si\,{\sc iv}/Si\,{\sc iii} absorbers along each sightline is given by:

%%%%%%%%%%%%%%%%%%%%%%%%%%%%%%%%%%%%%%%%%%%%%%%%%%%%%%%%

\begin{equation}
\Delta X\equiv (1+z)^2\,[\Omega_m(1+z)^2+\Omega_{\Lambda}]^{-1/2}\,\Delta z_{\rm abs}.
\end{equation}

%%%%%%%%%%%%%%%%%%%%%%%%%%%%%%%%%%%%%%%%%%%%%%%%%%%%%%%%

The integration over all sightlines then delivers $\Delta X_{\rm tot}$.
For our bias-corrected Si\,{\sc iv}/Si\,{\sc iii} absorber sample with 
log $N$(Si\,{\sc iii}$)\geq 12.2$ we have
$N$(H\,{\sc ii}$)_{\rm tot}\geq 5.4\times10^{20}$ cm$^{-2}$ and
$\Delta_{\rm tot}X=20.65$.
The cosmological mass density can then be written as 
$\Omega_b$(Si\,{\sc iii}$)\geq 4.6\times 10^{-4}\,Z_{\rm Si}$, 
where $Z_{\rm Si}$ is the Si abundance in the gas relative to 
the solar value, log (Si/H)$_{\sun}=-4.49$ (Asplund et al.\,2009).
Thus, if the CGM traced by Si\,{\sc iii} would have a solar Si abundance, 
it would contain roughly as much mass as the neutral ISM within galaxies at $z=0$ 
($\Omega_b$(H\,{\sc i}$)=4.7\pm 0.7 \times 10^{-4}$; Zwaan et al.\,2005).

On the one hand, $\Omega_b$(Si\,{\sc iii}$)$ could be much higher, if the mean 
metallicity of the gas was substantally lower than solar. This appears plausible
if most of the gas originates in the IGM, e.g., as material stemming 
from accretion flows. Lehner et al.\,(2013) studied the metallicity distribution
of LLS at $z\leq1$ and found two distinct populations
of absorbers with mean metallicities of $0.03$ solar (population\,I) and
$0.50$ solar (population\,II). They suggest that population\,I absorbers
represent metal-poor circumgalactic gas from accretion streams, while 
population\,II systems trace metal-enriched halo material from 
galaxy outflows and winds. In view of these results, we assume a value 
of $Z_{\rm Si}=0.5$ as a conservative upper limit for the metallicity of 
the Si\,{\sc iii} absorbers in our survey. This value leads to 
$\Omega_b$(Si\,{\sc iii}$)\geq 9.2\times 10^{-4}$, roughly twice the 
value of the neutral ISM within galaxies.

Earlier theoretical and observational studies that addressed 
the outflow of gaseous material from galaxies and the enrichment 
of the IGM suggested, on the other hand, that the metals produced within galaxies 
escape in the form of metal-rich (super-solar, eventually) gas 
pockets (Mac\,Low \& Ferrara 1999; Rauch et al.\,2001; Schaye, Carswell \& Kim 2007).
Such gas pockets could mimic the absorption properties of 
intervening metal absorbers, but they would carry only very little mass.
While it cannot be excluded that some of the Si\,{\sc iii} absorbers presented 
in this study belong to such metal patches with above-solar metallicities, the
large overall cross section of circumgalactic Si\,{\sc iii} together with the 
expected cosmological metal-mass density at $z=0$ (e.g., Calura \& Mateucci 2004) 
excludes that a dominating fraction of the circumgalactic Si\,{\sc iii} absorbers have 
metallicites above solar.

From the above considersations we conclude that the Si\,{\sc iv}/Si\,{\sc iii} 
bearing gas phase in the CGM of low redshift galaxies contains (possibly substantially)
more baryonic matter than their neutral ISM. This 
conclusion is supported by previous estimates of the baryon content 
of the cool, photoionized CGM in the local Universe from the 
COS-Halos survey (Werk et al.\,2014).

%%%%%%%%%%%%%%%%%%%%%%%%%%%%%%%%%%%%%%%%%%%%%%%%%%%%%%%%

\section{Discussion}

\subsection{On the origin of intervening Si\,{\sc iii} absorbers}

Our study indicates a tight spatial correlation between intervening
Si\,{\sc iii} absorbers at $z\leq 0.1$ and the local galaxy population
at impact parameters $\leq 400$ kpc, suggesting that the majority of 
these absorption systems arise in the extended gaseous halos of these galaxies.
This interpretation is supported by recent \emph{HST}/STIS and 
\emph{HST}/COS observations of Si\,{\sc iii} absorption in the CGM
of the Milky Way (Shull et al.\,2009; Collins et al.\,2009; 
Richter et al.\,2009; Lehner et al.\,2012; Herenz et al.\,2013; 
Richter et al.\,2016, in prep.), who show that doubly-ionized Si has the 
largest absorption cross section of all low, intermediate and high ions
with a sky covering fraction of $\langle f_c \rangle\approx 0.7$.
While the distances and the space distribution of Si\,{\sc iii} absorbers in the Milky Way 
halo still are unclear (owing to our internal vantage point), Lehner et al.\,(2015)
demonstrated that M31 also exhibits an extended, metal-enriched gaseous halo 
that gives rise to Si\,{\sc iii} absorption out to $\sim 200$ kpc.
Finally, other groups that have studied the absorption properties of the CGM for
individual galaxies (e.g., Keeney et al.\,2013), or pre-selected galaxy samples
such as the COS-Halos survey (e.g., Tumlinson et al.\,2011, 2013; Werk et al.\,2013;
Peeples et al.\,2013) demonstrate that Si\,{\sc iii} is ubiquitous in the
extended CGM of low-redshift galaxies. The existence of such discrete 
Si\,{\sc iii} bearing gas complexes with temperatures $T<10^5$ K
(thus below the virial temperature of their DM host halos) in the 
extended circumgalactic environment of galaxies can be understood in
terms of the steady circulation of hot and cold gas through the CGM
("circumgalactic fountain") as part of the ongoing galaxy evolution.
In this scenario, star formation in galaxies drives out large amounts of (hot) 
metal-enriched gas. Even if this material leaves the potential well
of an individual galaxy, it is trapped by the superordinate cosmological
structure.
From there, the gas may (slowly) fall back onto the galaxy 
of its origin in the form of
discrete gas structures, as the gas never reaches hydrostatic
equilibrium during such a circulation cycle (Ford et al.\,2014).
Alternatively, the gas may be accreted by a {\it different} galaxy
nearby, or heated up to the virial temperature of the superordinate 
DM halo, in which case it may remain extragalactic forever.

While the observed frequency of Si\,{\sc iii} absorbers with 
log $N$(Si\,{\sc iii}$)\geq 12.2$ and the derived impact parameter 
distribution are in line with a scenario, in which {\it all} Si\,{\sc iii}
absorbers are located within the virial radius of intervening galaxies
and thus are gravitationally bound to them, it cannot be excluded 
that there exists a population of intervening Si\,{\sc iii} absorbers that
traces gas at larger distances.
This is actually expected, since in many cases large, gas-rich spiral 
galaxies like the Milky Way do not represent isolated systems, but are 
part of a galaxy group, which binds metal-enriched 
diffuse gas within its own virial radius in the form of an intragroup medium.

Stocke et al.\,(2014) recently have studied a class of "warm" H\,{\sc i}/O\,{\sc vi} 
absorbers using COS spectra. They suggested that these systems trace an
extended (Mpc scale) $T=10^5-10^6$ K intragroup medium in spiral-rich galaxy groups.
If such warm gas was typical for group environments, it appears plausible 
that one would find cooler ($T<10^5$ K) gas patches embedded in such a medium that could
give rise to intervening Si\,{\sc iii}/Si\,{\sc iv} absorption. 
Some of the galaxy/absorber pairs that have apparent impact parameters in the
range $\rho=200-1000$ kpc may belong to such group absorbers.
The Cloudy modeling predicts that Si\,{\sc iii}/Si\,{\sc iv}
absorbers trace gas down to thermal pressures of $P/k=n_{\rm H}T\sim 1$ 
cm$^{-3}$\,K. For gas that is gravitationally bound to individual galaxies, 
such low gas pressures would be expected only in the outermost regions of 
galaxies near their virial radius (see also discussion in Shull 2014). 
However, such a value for $P/k$ also would be in agreement with the 
expected range of gas pressures in galaxy groups (Stocke et al.\,2013).

Our survey suggests that the number density distribution 
of Si\,{\sc iii} absorbers breaks down for column densities 
log $N$(Si\,{\sc iii}$)\leq 12.2$ (Fig.\,4, right panel), which
is not a completeness effect in our data (see Sect.\,4.1). These low-column
density systems may represent the prime candidates for metal-enriched 
cloudlets that arise in regions with low gas densities
(and pressures), such as in group environments and in the IGM. 
As pointed out by Stocke et al. (2013), the expected number density of 
galaxy groups is ten times less than the space density of $L^{\star}$
galaxies (Berlind et al.\,2006).
Interestingly, the number density of Si\,{\sc iii} absorbers with 
log $N$(Si\,{\sc iii}$)\leq 12.2$ is also $\sim 10$ times less than
the number density of absorbers with log $N$(Si\,{\sc iii}$)\geq 12.2$,
supporting a scenario in which high-column density Si\,{\sc iii} systems
trace the CGM of galaxies, while low-column density Si\,{\sc iii} systems
arise in the intragroup gas of galaxy groups.

%%%%%%%%%%%%%%%%%%%%%% FIGURE 16 %%%%%%%%%%%%%%%%%%%%%%

\begin{figure*}[t!]
\resizebox{1.0\hsize}{!}{\includegraphics{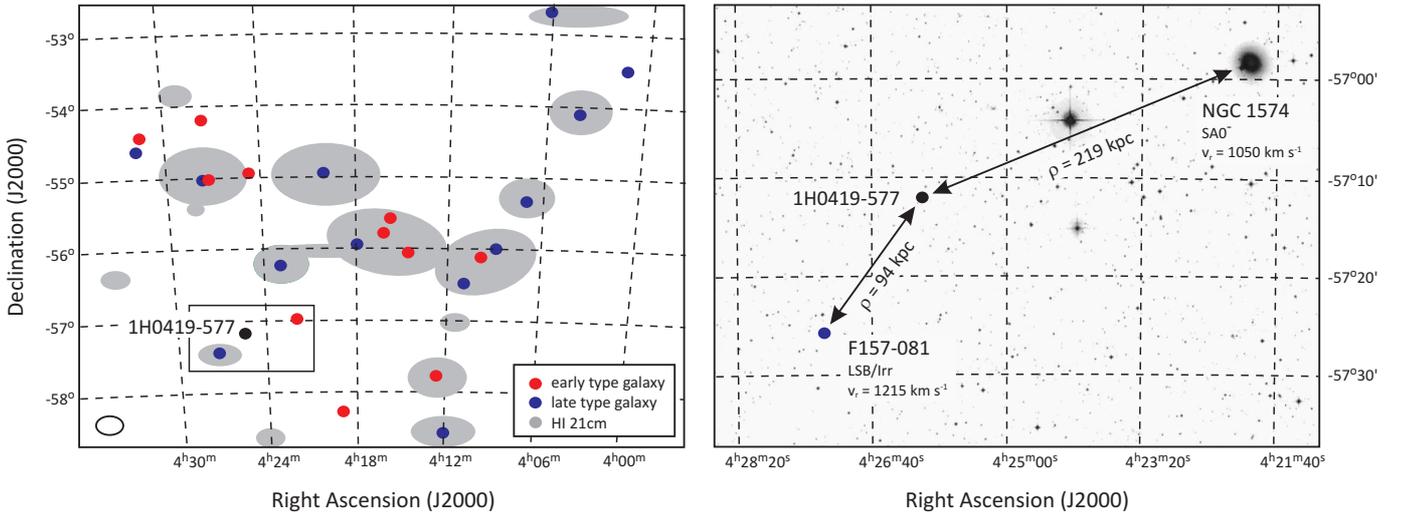}}
\caption[Dorado galaaxy group]{
{\it Left panel:} Distribution of galaxies and H\,{\sc i} in the Dorado galaxy
group (see Kilborn et al.\,2005 and references therein).
Intervening Si\,{\sc iii} absorption is observed at $z=0.0037$ ($cz=1109$ km\,s$^{-1}$)
towards the QSO 1H4019$-$577 (black filled circle). The absorption most likely
is associated with the CGM of one group member or with the intergroup
medium. {\it Right panel:} Zoom-in for the region around the two group
members F157$-$081 and NGC\,1574; the absorber could be associated with
either or both of these galaxies given its velocity and projected distance
to each of the systems.
}
\end{figure*}

%%%%%%%%%%%%%%%%%%%%%%%%%%%%%%%%%%%%%%%%%%%%%%%%%%%%%%%%

With the occurrence of several galaxies at impact parameters 
$\rho < 1000$ kpc to the QSO sightlines and with radial velocities close 
to the observed absorber velocities (such as expected for group environments),
the interpretation of the observed absorption with respect to origin and nature 
of the gas in its galaxy environment generally is tricky.
While a more systematic investigation between intervening Si\,{\sc iii}
absorbers and galaxy group environments clearly is beyond the scope of this 
paper, we show in Fig.\,16 as a prominent example the galaxy group environment 
of the $z=0.0037$ Si\,{\sc iii} absorber towards the QSO 1H0419$-$577
(IRAS\,F04250$-$5718).
This sightline passes two galaxies in the Dorado galaxy group
(Kilborn et al.\,2005; Maia, Da Costa \& Latham 1989) 
at impact parameters $\rho<220$ kpc. In principle, the observed 
Si\,{\sc iii} absorption at $cz=1109$ km\,s$^{-1}$ could be associated with either of these
two galaxies, which have radial velocities of $v_{\rm r}=1050$ and $1215$ km\,s$^{-1}$
(Kilborn et al.\,2005). The observed gas may be infalling or outflowing gaseous material; 
the metallicity of the absorber lies in the range between $0.1-1.0$ solar 
(as derived from the fit of the H\,{\sc i} Ly\,$\alpha$ absorption together
with the Cloudy model of the Si\,{\sc iii}/Si\,{\sc iv} absorption), thus in
line with both scenarios. However, the absorber could also be located outside the
virial radii of these two galaxies and may belong to a faint (unseen) 
dwarf galaxy in the same group or may represent a metal-rich gas patch 
embedded in the intragroup medium of the Dorado group. The interpretation 
of {\it individual} absorber/galaxy pairs thus remains inconclusive with
magnitude-limited galaxy data. The statistical connection between
galaxies and intervening Si\,{\sc iii} systems, as studied here, yet demonstrates
that both class of objects trace the same Mpc-scale environment within the cosmic 
web.

\subsection{Comparison with previous Si\,{\sc iii} absorption-line studies}

We are not aware of any other systematic studies in the literature that focus
explicitly on Si\,{\sc iii}-selected intervening absorption systems and their relation
to galaxies. There are, however, several absorption-line surveys at low redshift
(using various UV spectrographs) that provide information on the 
number density of Si\,{\sc iii} and other intermediate ions (e.g., C\,{\sc iii})
and the distribution of equivalent widths/column densities of these ions.

Tilton et al.\,(2012) compiled UV absorption-line data from \emph{HST}/STIS and
\emph{FUSE} and prepared a catalog of UV absorbers for redshifts $z<0.4$. From
their data sample they derive a number density of intervening Si\,{\sc iii} 
absorbers of $d{\cal N}/dz=7.4^{+2.1}_{-1.2}$ for log $N$(Si\,{\sc iii}$)>12.2$
for this redshift range, based on a total redshift path of $\Delta z\approx 5.2$.
This value is more than twice the value derived in this study.
The same authors recently published another absorption-line catalog of intervening
absorption systems at $z<0.4$, this time based on \emph{HST}/COS data of
75 QSOs (Danforth et al.\,2016). In this new survey, the authors derive a 
number density of $d{\cal N}/dz\approx7$ for log $N$(Si\,{\sc iii}$)>12.2$
and $z<0.1$. Moreover, their study suggests an {\it increase} for 
$d{\cal N}/dz$(Si\,{\sc iii}) for decreasing redshift in the local
Universe (their Fig.\,15). Their new value again is substantially higher than the
value derived by us. To investigate the origin for this discrepancy we have
carefully compared their absorber list with ours, as our COS data sample includes all of 
the 75 QSOs presented in Danforth et al.\,(2016). We suspect that the higher value
for $d{\cal N}/dz$(Si\,{\sc iii}) derived by Danforth et al.\,(2016) stems 
from the less stringent selection criteria for identifying metal absorbers and 
defining the relevant absorption path lengths along their sightlines. 
Also, a possibly existing selection bias in their QSO sample may be
the reason for the higher value of $d{\cal N}/dz$(Si\,{\sc iii}).

We identify three Si\,{\sc iii} systems in their absorber list, 
whose identifications are based on absorption features that (in our opinion) most 
likely have different origins.
We also have identified one candidate Si\,{\sc iii} absorber 
that is not listed in the Danforth et al.\,(2016) paper. 
In Table A.12 in the Appendix we list these discrepant absorption systems together
with a short description of the absorption characteristics.

As we discuss in one the following sections, state-of-the art
hydrodynamical simulations of metal-enriched gas in the local Universe
do not support values of $d{\cal N}/dz>5$ for Si\,{\sc iii} absorbers
for log $N$(Si\,{\sc iii}$)>12.2$, but favour Si\,{\sc iii} number 
densities of $<4$ for $z\approx0$.

\subsection{Comparison with other CGM studies}

As mentioned above, Si\,{\sc iii} is frequently detected in the CGM of
individual galaxies (Keeney et al.\,2013; Lehner et al.\,2015) and in
the CGM of pre-selected galaxy samples (e.g., COS-Halos; Tumlinson et al.\,2013;
Werk et al.\,2013; Liang \& Chen 2014). Our study, which represents a statistical rather than
a targeted study of the CGM in the local Universe, complements most
of the previous observational CGM studies. 
Many of the COS sightlines that have been used by previous, targeted
CGM studies are included in our sample. Because of the statistical nature
of our survey, we refrain from comparing our results for {\it individual} 
galaxy/absorber pairs with results from other studies.

From an archival study of CGM absorption around 195 galaxies at $z<0.176$
Liang \& Chen (2014) find a mean Si\,{\sc iii} covering fraction of $0.60^{+0.13}_{-0.18}$
within $0.54\,R_{\rm halo}$ and $0.14^{+0.11}_{-0.05}$ for the range $0.54-1.02\,R_{\rm halo}$,
but no significant Si\,{\sc iii} absorption beyond. These covering fractions
are lower than our estimate of $\langle f_c \rangle=0.69\pm0.11$. This is not surprising,
however, since 
their study considers only strong Si\,{\sc iii} systems with $W_{\rm r}>100$ m\AA\,
(log $N$(Si\,{\sc iii}$)>12.6$), while our estimate is based on a more sensitive search 
including weak and strong absorbers
with log $N$(Si\,{\sc iii}$)>12.2$. If we adopt our measured value of 
$d{\cal N}/dz$(Si\,{\sc iii}$)=1.9$ for $N$(Si\,{\sc iii}$)\geq 12.6$ for
our estimate of $\langle f_c \rangle$ using equation (4), we obtain a value
of $0.47^{+0.09}_{-0.08}$ within $R_{\rm vir}$, thus in good agreement with the Liang \& Chen (2014) 
estimate. Also Werk et al.\,(2013) consider only strong Si\,{\sc iii} absorbers
around $L\approx L^{\star}$ absorbers with $W_{\rm r}>100$ m\AA\, and find
$\langle f_c \rangle=0.72^{+0.07}_{-0.17}$ for their total galaxy with little 
dependence on the galaxy mass or star-formation rate (their Table 6). Within
the given error bars, these estimates for $\langle f_c \rangle$ (based on
different methods) agree very well with each other. 

In contrast, Bordoloi et al.\,(2014) studied the CGM around {\it dwarf} 
galaxies at $z\leq 0.1$ and found no metal absorption beyond $0.5R_{\rm vir}$.
Their study suggests that the filling factor of warm circumgalactic gas is
substantially smaller for dwarfs than for $L^{\star}$ galaxies.
As described in Sect.\,4.2, our full data set includes most of the sightlines from 
the Bordoloi et al.\,sample, so that we can analyze the Si\,{\sc iii} 
absorption fraction in the CGM of these dwarf galaxies directly. 
We find that only 17 out of the according 40 sightlines show detectable Si\,{\sc iii} 
with log $N$(Si\,{\sc iii}$)>12.2$ at the redshifts of the dwarf 
galaxies ($\langle f_c \rangle=0.43^{+0.10}_{-0.11}$), demonstrating
that the absorption fraction of the warm CGM traced by Si\,{\sc iii} around dwarf 
galaxies is only $\sim 60$ percent of that seen in galaxies with higher luminosities. 
The small CGM absorption fraction around dwarf galaxies is the
reason for the almost identical number densities of intervening Si\,{\sc iii}
in our original data sample and in the bias-corrected sample (Sect.\,2.4).

Turning to high-ionization species,
Wakker \& Savage (2009) have studied the relation between O\,{\sc vi}
and nearby galaxies at $z<0.017$ based on UV and FUV data along 
76 extragalactic sightlines, demonstrating that a substantial fraction of the
O\,{\sc vi} systems arise from gas within $R_{\rm vir}$ of these galaxies
(see also Prochaska et al.\,2011; Tumlinson et al.\,2011; Stocke et al.\,2013).
However, in contrast to Si\,{\sc iii}, O\,{\sc vi} is also present beyond
the virial radius of these galaxies, tracing hot metal-enriched gas (at higher
gas temperatures and lower densities than Si\,{\sc iii})
that presumably has been ejected by galactic winds and outflows in past 
epochs of active star formation. In fact, in view of our CGM modeling results
presented in Sect.\,5, the measured O\,{\sc vi} number density at low redshifts of
$d{\cal N}/dz$(O\,{\sc vi}$)\approx 16$ (Tripp et al.\,2008) indicates a very
large absorption cross section of O\,{\sc vi} that is clearly beyond the expected
cross section of the CGM in the local Universe, even if non-standard
Schechter parameters in the luminosity function are considered (Table 1, fourth row).
The Milky Way possibly also contains
a massive, extended O\,{\sc vi} envelope that reaches deep into
the Local Group potential well (Sembach et al.\,2003 and Wakker et al.\,2003).

Interestingly, the detailed analysis of the
COS-Halos sample suggests that the absorption properties of low and 
intermediate ions (such as Si\,{\sc iii}) in star-forming galaxies 
does not significantly differ from those in passive galaxies, while
O\,{\sc vi} predominantly arises in the halo of active galaxies
(Thom et al.\,2012; Werk et al.\,2013; Tumlinson et al.\,2011).
From their study, Liang \& Chen (2014) conclude that the CGM becomes 
progressively more ionized from small to large radii. A similar 
conclusion was drawn by us from modeling the radial decline of the
neutral gas fraction in the CGM in the Local Group (Richter 2012).

%%%%%%%%%%%%%%%%%%%%%% FIGURE 17 %%%%%%%%%%%%%%%%%%%%%%

\begin{figure}[t!]
\resizebox{0.9\hsize}{!}{\includegraphics{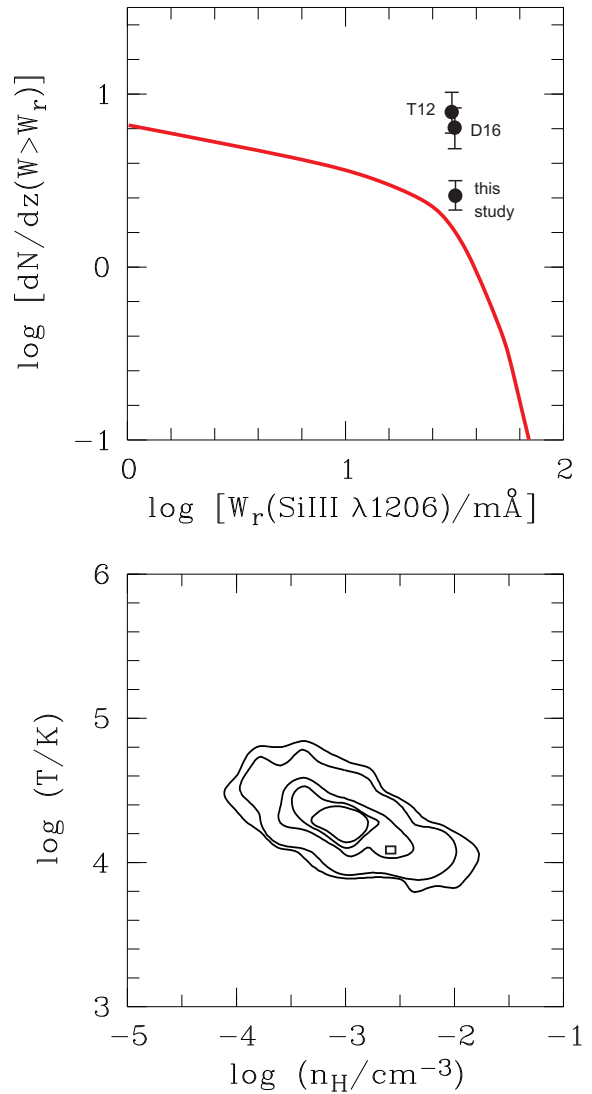}}
\caption[Dorado galaaxy group]{
{\it Upper panel:} Cumulative number density of intervening Si\,{\sc iii}
absorbers for different equivalent-width limits predicted by a simulation run from
the OverWhelmingly Large Simulations (OWLS; Schaye et al.\,2010), as presented in
Tepper-Garc{\'{\i}}a et al.\,(2012). The measured value from this study and from
Tilton et al.\,(2012; T12) and Danforth et al.\,(2016; D16) are indicated with
filled circles.
{\it Lower panel:} Phase diagram of intervening Si\,{\sc iii} absorbers
in the temperature-density plane, as predicted by OWLS.
}
\end{figure}

%%%%%%%%%%%%%%%%%%%%%%%%%%%%%%%%%%%%%%%%%%%%%%%%%%%%%%%%

\subsection{Comparison with hydrodynamical simulations}

Cosmological, hydrodynamical simulations have recently become 
a powerful tool for studying intergalactic and circumgalactic gas at low and high
redshifts (e.g., Oppenheimer \& Dav{\'e} 2006; Crain
et al.\,2013; Nuza et al.\,2014). The line statistics from synthetic absorption 
spectra generated from such simulations can be directly compared to QSO 
absorption-line studies to investigate the different gas phases in the
IGM and CGM and their redshift evolution (Fang \& Bryan\,2001; 
Richter et al.\,2006; Fangano, Ferrara \& Richter 2007; 
Tepper-Garc{\'{\i}}a et al.\,2011; 2012; 2013; Oppenheimer et al.\,2012; 
Smith et al.\,2011; Churchill et al.\,2015).

While most of these studies have focused on the properties of high-ion
absorbers, such as O\,{\sc vi} and Ne\,{\sc viii}, to study the shock-heated
warm-hot intergalactic medium or shock-heated gas in the CGM,
intermediate ions such as Si\,{\sc iii} and C\,{\sc iii} so far have not been 
considered for a detailed comparison study of absorption-line statistics in 
simulations and observational data. From the simulations of Smith et al.\,(2011),
Danforth et al.\,(2016) extracted the cumulative line density of intervening
metal absorbers for a comparison with their COS absorption-line survey. 
While their own value for $d{\cal N}/dz$(Si\,{\sc iii}) of $\sim 7$ 
for log $N$(Si\,{\sc iii}$)>12.2$ (see Sect.\,8.2) lies far above the value 
predicted by Smith et al. for this column-density limit 
($d{\cal N}/dz$(Si\,{\sc iii}$)\approx3$), the Si\,{\sc iii} number density 
derived by us is in very good agreement with the results by Smith et al.

We here re-use synthetic spectra generated along random LOS from a simulation
run (Model AGN) of the OverWhelmingly Large Simulations (OWLS; Schaye et al.\,2010)
to further investigate the frequency and physical conditions in Si\,{\sc iii} absorbers
at low redshift ($z=0.25$). Similar simulations have been used by us previously
to study absorption signatures of warm-hot gas in the local Universe
(Tepper-Garc{\'{\i}}a et al.\,2011, 2012, 2013).
The OWLS simulations were set up using the smoothed particle hydrodynamics (SPH)
code {\sc GADGET III} (a modified version of {\sc GADGET II}; Springel 2005) and
were carried in a cubic box of $100\,h^{-1}$ comoving
Mpc on a side, containing 5123 dark matter (DM) and 5123 baryonic particles.
The initial mass resolution is $4.1\times 10^8\,h^{-1}\,M_{\sun}$ (DM) and 
$8.7\times 10^8\,h^{-1}\,M_{\sun}$ (baryonic), while the
gravitational softening is set to $8\, h^{-1}$ comoving kpc and is fixed at
$2\,h^1{-1}$ proper kpc below $z=3$.
The cosmological simulations from the OWLS project are characterized by a particularly careful
implementation of important physical processes that have been largely ignored 
in earlier studies (e.g., the influence of the photoionization on the cooling function 
of the gas).
Details on the sub-grid physics used in OWLS can be found in Wiersma et al.\,(2009a; radiative cooling), 
Schaye \& Dalla Vecchia (2008; star formation), Dalla Vecchia \& Schaye (2008; stellar feedback), and
Booth \& Schaye (2009; AGN feedback).
The strategy, how the synthetic absorption-line spectra from OWLS were set up and analyzed
is presented in Tepper-Garc{\'{\i}}a et al.\,(2011,2012,2013).
In Fig.\,17, upper panel, we show the cumulative number density of intervening Si\,{\sc iii}
absorbers as a function of the limiting equivalent width predicted by OWLS (Model AGN).
The redshift in the simulation run is higher than the maximum redshift of our data. 
This is, however, not an issue since we do not expect the gas producing the observed 
absorption to evolve strongly from $z=0.25$ to $z=0.10$.
The measured value from this study and from Tilton et al.\,(2012; T12) and Danforth 
et al.\,(2016; D16) are indicated with filled circles. The predicted number
density is $\sim 1.7$ for log $N$(Si\,{\sc iii}$)>12.2$ 
(log [$W_{\rm r}(\lambda1206)$/m\AA$]\geq 1.52)$, thus only slightly 
below the observed value of $2.5\pm 0.4$ from our COS data set. 

It is worth noting that differences between the predicted and observed 
line-number densities may be alleviated considering that the stellar yields used 
in the simulations to model the chemical evolution of the gas are uncertain by 
factors of a few (see Wiersma et al.\,2009b). 
Thus, some modification of the abundances in post-processing would be justified, 
which would bring the predicted $d{\cal N}/dz$ into better agreement with our measured value. 
However, we refrain from doing so because this could break the self-consistency of the simulation 
if it were to change the cooling rates significantly.

In the lower panel of Fig.\,17 we show the phase-space distribution 
in the temperature-density plane of the Si\,{\sc iii}-absorbing gas in the AGN run at
$z=0.25$ from OWLS. The 
typical Si\,{\sc iii} absorber in the simulation has a temperature of $T\approx 15,000-20,000$ K
and a density of $n_{\rm H}\approx 10^{-4}-10^{-2}$ cm$^{-3}$, in very good agreement with
the values estimated with Cloudy from the Si\,{\sc ii}/Si\,{\sc iii}/Si\,{\sc iv} column-density
ratios in our COS absorber sample. The resulting range for the expected thermal gas pressure
is $P/k\approx 2-160$ cm$^{-3}$\,K, thus in a range that is typical for the CGM (Cen 2013).

In conclusion, the AGN model from OWLS and the simulations from Smith et al.\,(2011) predict 
absorber properties that are compatible with our observations and in line with
the idea, that the majority of the Si\,{\sc iii} absorbers at $z\leq 0.1$ arise in the 
extended halos of low-redshift galaxies. For the future, we are planning
to further explore the relation between intermediate-ion absorbers and galaxies at low $z$
using higher-resolution simulations from the EAGLE (Evolution and Assembly of GaLaxies 
and their Environments) project (Schaye et al.\,2015).

%%%%%%%%%%%%%%%%%%%%%%%%%%%%%%%%%%%%%%%%%%%%%%%%%%%%%%%%

\section{Summary and Conclusions}

In this paper, we present a systematic study of intervening
Si\,{\sc iii} absorbers and their relation to galaxies 
in the redshift range $z\leq0.1$ along 303 QSO sightlines 
using archival UV spectra obtained with \emph{HST}/COS. 
The main results are summarized as follows:\\
\\
(1) We detect 69 intervening Si\,{\sc iii} absorbers in the range
$z=0.00014-0.09789$ along a total
redshift path of $\Delta z \approx 24$. The restframe equivalent widths in the
Si\,{\sc iii} $\lambda 1206$ line lie in the range $W_{\rm r}=13-885$ m\AA,
corresponding to Si\,{\sc iii} column densities log $N$(S\,{\sc iii}$)>11.9$.
We derive a bias-corrected number density of intervening Si\,{\sc iii} systems
of $d{\cal N}/dz=2.5\pm 0.4$ for absorbers with log $N$(Si\,{\sc iii}$)\geq12.2$,
suggesting that Si\,{\sc iii}-selected absorbers outnumber strong 
Mg\,{\sc ii} at $z\approx0$ by a factor of $\sim 3$. Weak absorbers in the range
log $N$(Si\,{\sc iii}$)=11.8-12.2$ turn out to be rare ($d{\cal N}/dz\approx0.3$);
they possibly represent an absorber population that is distinct from the one with
larger Si\,{\sc iii} column densities.\\
\\
(2) The observed Si\,{\sc iii} absorption is typically accompanied by absorption 
from other low and/or high ions. The positive correlations between 
the equivalent widths of Si\,{\sc iii} and Si\,{\sc iv} and between
Si\,{\sc iii} and Si\,{\sc ii} indicates that doubly ionized silicon
traces gas in a range of physical conditions. This 
conclusion is supported by Cloudy ionization models of the 
absorbers, which indicate that Si\,{\sc iii} arises either in diffuse, ionized
low-density (log $n_{\rm H}\leq -3$, typically) gas that is also observed in
Si\,{\sc iv} absorption, or in somewhat denser (log $n_{\rm H}\geq -3$, typically),
more neutral environments that are also traced by Si\,{\sc ii}.\\
\\
(3) We geometrically model the absorption cross section of diffuse gas within the
virial radii ($R_{\rm vir}$) of galaxies at $z=0$ for galaxies in the luminosity
range $L/L^{\star}>0.001$ using the SDSS galaxy luminosity function for $z=0$
and scaling relations between $R_{\rm vir}$ and $L$. 
Our modeling implies that circumgalactic gas within the virial radius of 
galaxies contributes with $d{\cal N}/dz<9$ to the total number density
of intervening absorbers, even if a unity covering fraction
of the absorbing gas is assumed. If we adopt the SDSS $g$-band luminosity
function as reference (Montero-Dortas et al.\,2009), we obtain $d{\cal N}/dz=3.6$
as expected number density for CGM absorbers from $L\geq0.001\,L^{\star}$ galaxies.
In light of this result, the measured value
of $d{\cal N}/dz$(Si\,{\sc iii}$)=2.5\pm0.4$ is in line with the idea, that the
majority of the intervening Si\,{\sc iii} absorbers with
log $N$(Si\,{\sc iii}$)\geq12.2$ at $z\approx0$ trace the 
circumgalactic medium at $R<R_{\rm vir}$ of nearby galaxies with an 
average covering fraction of $\langle f_c\rangle=2.5/3.6=0.69$,
a value that is in excellent agreement with results from other 
studies (Liang \& Chen 2014; Werk et al.\,2013). 
Our model further predicts that intervening metal absorbers with
number densities $d{\cal N}/dz>9$ (e.g., O\,{\sc vi} absorbers) 
must partially arise from gas outside of the virial radius of galaxies.\\
\\
(4) We compare the Si\,{\sc iii} absorber redshifts and positions with that of
$\sim 64,000$ galaxies at $z\leq0.1$ using archival galaxy-survey data.
For $\sim 60$ percent of the absorbers we find possible host galaxies
with $L\geq 0.01\,L^{\star}$
within $300$ km\,s$^{-1}$ of the absorbers and at impact parameters
$\rho<200$ kpc, indicating that the spatial distributions of
Si\,{\sc iii} absorbers and galaxies are highly correlated. We verify
the significance of this result by studying the distribution of galaxies 
around the same sightlines in velocity ranges without Si\,{\sc iii} absorption
and find no clustering of the galaxies, as expected. 
The observed Si\,{\sc iii} absorption fraction around the galaxies scales
differently with impact parameter than predicted by our model. We assign
this behavior to the incompleteness in our galaxy sample to identify
faint galaxies that are associated with Si\,{\sc iii} 
absorbers.\\
\\
(5) We estimate the baryon content of the Si\,{\sc iii}-absorbing gas phase
in the CGM at low $z$ using Cloudy ionization models. Assuming that 
the metallicity of the gas is $\leq 0.5$ solar, the cosmological mass density 
of the gas in Si\,{\sc iii} absorbers is 
$\Omega_b$(Si\,{\sc iii}$)\geq 9.2\times 10^{-4}$. 
This lower limit is roughly twice the value derived for the H\,{\sc i} mass 
density at $z=0$, indicating that the diffuse CGM around galaxies, as traced by 
doubly-ionized Si, contains substantially more baryonic matter than
their neutral ISM. \\
\\
(6) We discuss the origin of intervening Si\,{\sc iii} absorption in the CGM of galaxies
and in galaxy groups and compare our results with previous absorption-line 
catalogs and CGM studies at low redshift. We suggest that the majority of
the strong Si\,{\sc iii} absorbers with log $N$(Si\,{\sc iii}$)\geq12.2$
typically arise within the extended halos of galaxies within $R_{\rm vir}$, 
while weak systems with log $N$(Si\,{\sc iii}$)<12.2$ possibly also 
trace gas at larger distances (e.g., group environments or the IGM). 
The comparison of our measurements with predictions from
cosmological hydrodynamical simulations from the OWLS project and other simulations 
shows a fair agreement for the absorber number density and the estimated
temperature-density range of intervening Si\,{\sc iii} absorbers.\\

%%%%%%%%%%%%%%%%%%%%%%%%%%%%%%%%%%%%%%%%%%%%%%%%%%%%%%%%

\begin{acknowledgements}

This research used the facilities of the Canadian Astronomy Data Centre
operated by the National Research Council of Canada with the support of
the Canadian Space Agency.
This research also has made use of the SIMBAD database,
operated at CDS, Strasbourg, France. We thank Joop Schaye and 
an anonymous referee for helpful comments and suggestions.

\end{acknowledgements}

%%%%%%%%%%%%%%%%%%%%%%%%%%%%%%%%%%%%%%%%%%%%%%%%%%%%%%%%

%\begin{thebibliography}{}
\section*{References}

\begin{small}
\begin{noindent}
Anderson, M.E., \& Bregman, J.N. 2011, ApJ, 737, 22
\\
Asplund, M., Grevesse, N., Sauval, A.J., \& Scott, P. 2009, ARA\&A, 47, 481
\\
Berlind, A.A., Frieman, J., Weinberg, D.H. 2006, ApJS, 167,  2006, ApJS 167, 1
\\
Birnboim, Y.\& Dekel, A. 2003, MNRAS, 345, 349
\\
Bogd{\'a}n, {\'A}, Forman, W.R., Kraft, R.P., \& Jones, C. 2013, ApJ, 772, 98
\\
Burchett, J.N., Tripp, T.M., Werk, J.K., et al. 2015, ApJ, 815, 91
\\
Calura, F. \& Mateucci, F. 2004, MNRAS, 350, 351
\\
Cen, R. 2013, ApJ, 770, 139
\\
Churchill, C.W., Van\,der Vliet, J.R., Trujillo-Gomez, S., Kacprzak, G.G., \& Klypin, A. 2015, ApJ, 802, 10
\\
Collins, J.A., Shull, J.M., \& Giroux, M.L. 2009, ApJ, 705, 962
\\
Crain, R.A., McCarthy, I.G., Schaye, J., Theuns, T., \& Frenk, C.S. 2013, MNRAS, 432, 3005
\\
Dai, X., Anderson, M.E., Bregman, J.N., \& Miller, J.M. 2012, ApJ, 755, 107
\\
Danforth, C.W., Tilton, E.M., Shull, J.M. et al.\,2016, ApJ, 817, 111
\\
Debes, J., et al. 2016, Cosmic Origins Spectrograph Instrument Handbook,
Version 8.0 (Baltimore, MD: STScI)
\\
Fang, T., \&  Bryan, G.L. 2001, ApJ, 561, 31
\\
Fangano, A.P.M., Ferrara, A., \& Richter, P. 2007, MNRAS, 381, 469
\\
Ferland, G.J., Korista, K.T., Verner, D.A., Fergueson, J.W.,
Kingdon, J.B., \& Verner, E.M. 1998, PASP, 110, 761
\\
Fontana, A., \& Ballester, P. 1995, ESO Messenger, 80, 37
\\
Ford, A. B., Dav{\'e}, R., Oppenheimer, B. D., et al. 2014, MNRAS, 444, 1260
\\
Fox, A.J., Wakker, B.P.; Savage, B.D., et al. 2005, ApJ, 630, 332
\\
Fox, A.J., Savage, B.D., \& Wakker, B.P. 2006, ApJS, 165, 229
\\
Fox, A.J., Wakker, B.P., Smoker, J.V., et al. 2010, ApJ, 718, 1046
\\
Green, J.C., Froning, C.S., Osterman, S., et al. 2012, ApJ, 744, 60
\\
Haardt, F., \& Madau, P. 2012, ApJ, 746, 125
\\
Herenz, P., Richter, P., Charlton, J.C., \& Masiero, J.R. 2013, A\&A, 550, A87
\\
Hinshaw, G., Larson, D., Komatsu, E., et al.\,2013, ApJS, 208, 19
\\
Kacprzak, G.G., Churchill, C.W., Steidel, C.C., \& Murphy, M.T. 2008, ApJ, 135, 922
\\
Kalberla, P.M.W., Burton, W.B., Hartmann, D., et al.\,2005, A\&A, 440, 775
\\
Keeney, B.A., Stocke, J.T., Rosenberg, J.L., et al.\,2013, ApJ, 765, 27
\\
Kere$\check{s}$, D., Katz, N., Weinberg, D.H., \& Dav{\'e}, R. 2005, MNRAS, 363, 2
\\
Kilborn, V.A., Koribalski, B., Forbes, D.A., Barnes,D.G., \& Musgrave, R.C. 2005, MNRAS, 356, 77
\\
Lehner, N., Howk, J.C., Thom, C., et al. 2012, MNRAS, 424, 2896
\\
Lehner, N., Howk, J.C., Tripp, T.M., et al. 2013, ApJ, 770, 138
\\
Lehner, N., Howk, J.C., \& Wakker, B.P. 2015, ApJ, 804, 79
\\
Liang, C.J., \& Chen, H.-W. 2014, MNRAS, 445, 2061
\\
Lundgren, B.F., Brunner, R.J., York, D.G., et al. 2009, ApJ, 698, 819
\\
Mac\,Low, M.M., \& Ferrara, A. 1999, ApJ, 513, 142
\\
Maia, M.A.G., Da Costa, L.N., \& Latham, D.W. 1989, ApJS, 69, 809
\\
Maller, A.H., \& Bullock, J.S. 2004, MNRAS, 355, 694
\\
Mo, H.J., \& Miralda-Escude, J. 1996, ApJ, 469, 589
\\
Montero-Dorta, A.D., \& Prada, F. 2009, 399, 1106
\\
Morton, D.C. 2003, ApJS, 149, 205
\\
Moster, B.P., Somerville, R.S., Maulbetsch, C., et al.\,2010, ApJ, 710, 903
\\
Nestor, D.B., Turnshek, D.A., \& Rao 2005, ApJ, 628, 637
\\
Nielsen, N.M., Churchill, C.W., \& Kacprzak, G.G. 2013, ApJ, 776, 115
\\
Nuza, S.E., Parisi, F., \& Scannapieco, C.2014, MNRAS, 441, 2593
\\
Oppenheimer, B., \& Dav{\'e}, R. 2006, MNRAS, 373, 1265
\\
Oppenheimer, B., Dav{\'e}, R., Katz, N., Kollmeier, J.A., \& Weinberg, D.H. 2012, MNRAS, 420, 829
\\
Peacock, J.A., \& Smith, R.E. 2000, MNRAS, 318, 1144
\\
Peeples, M.S., Werk, J.K., Tumlinson, J., et al.\,2014, ApJ, 786, 54
\\
Planck; Fermi Collaborations, et al.\,2015, arXiv: 1502.01589
\\
Prochaska, J.X., Weiner, B., Chen, H.-W., Mulchaey, J., \& Cooksey, K.
2011, ApJ, 740, 91
\\
Prochter, G.E., Prochaska, J.X., Chen, H.-W., et al. 2006, ApJ, 639, 766
\\
Putman, M.E., Peek, J.E.G., \& Joung, M.R. 2012, ARA\&A, 50, 491
\\
Rauch, M., Sargent, W.L.W., Barlow, T.A., \& Carswell, R.F. 2001, ApJ, 562, 76
\\
Rees, M.J., Ostriker, J.P., 1977, MNRAS, 179, 541
\\
Richter, P., Savage, B.D., Wakker, B.P., Sembach, K.R., Kalberla, P.M.W.
2001, ApJ, 549, 281
\\
Richter, P., Fang, T., \& Bryan, G. L. 2006, A\&A, 451, 767
\\
Richter, P. 2006, Reviews in Modern Astronomy 19, 31
\\
Richter, P., Charlton, J.C., Fangano, A.P.M., Ben Bekhti, N.,
\& Masiero, J.R. 2009, ApJ, 695, 1631
\\
Richter, P., Krause, F., Fechner, C., Charlton, J.C., 
\& Murphy, M.T. 2011, A\&A, 528, A12
\\
Savage, B.D., \& Sembach, K.R. 1991, ApJ, 379, 245
\\
Schaye, J. Carswell, R.F., \& Kim, T.-S. 2007, MNRAS, 379, 1169
\\
Schaye, J., Dalla Vecchia, C., Booth, C.M., et al. 2010, MNRAS, 402, 1536
\\
Schaye, J., Crain, R.A., Bower, R.G., et al. 2015, MNRAS, 446, 521
\\
Schechter, P. 1976, ApJ, 203, 297
\\
Sembach, K.R., Wakker, B.P., Savage, B.D., Richter, P. et al. 2003, ApJS, 146, 165
\\
Shull, J.M., Jones, J.R., Danforth, C.W., \& Collins, J.A. 2009, 699, 754
\\
Shull, J.M. 2014, ApJ, 784, 142
\\
Smith, B.D., Hallman, E.J., Shull, J.M., O'Shea, B.W. 2011, ApJ, 731, 6
\\
Stocke, J.T., Keeney, B.A., Danforth, C.W., et al.\,2013, ApJ, 763, 148
\\
Stocke, J.T., Keeney, B.A., Danforth, C.W., et al.\,2014, ApJ, 791, 128
\\
Tepper-Garc{\'{\i}}a, T., Richter, P., Schaye, J., et al. 2011, MNRAS, 413, 190
\\
Tepper-Garc{\'{\i}}a, T., Richter, P., Schaye, J., et al. 2012, MNRAS, 425, 1640
\\
Tepper-Garc{\'{\i}}a, T., Richter, P., Schaye, J., et al. 2013, MNRAS, 436, 2063
\\
Thom, C., Tumlinson, J.,Werk, J., Prochaska, J.X., \& Tripp, T. 2012, ApJ, 758, 41
\\
Tilton, E.M., Danforth, C.W., Shull, J.M., \& Ross, T.L. 2012, ApJ, 759, 112
\\
Tripp, T.M., Wakker, B.P., Jenkins, E.B., et al. 2003, AJ, 125, 3122
\\
Tripp, T.M., Sembach, K.R., Bowen, D.V., et al. 2008, ApJS, 177, 39
\\
Tumlinson, J., Shull, J.M., Rachford, B.L., et al.\,2002, ApJ, 566, 857
\\
Tumlinson, J., Thom, C., Werk, J.K., et al.\,2011, Sci, 334, 948
\\
Tumlinson, J., Thom, C., Werk, J.K., et al.\,2013, ApJ, 777, 59
\\
van de Voort, F., Schaye, J., Booth, C.M., \& Dalla Vecchia, C. 2011, MNRAS, 414, 2458
\\
Wakker, B.P. \& van\,Woerden, H. 1998, ARA\&A, 35, 217
\\
Wakker, B.P., Howk, J.C., Savage, B.D. 1999, Nature, 402, 388
\\
Wakker, B.P. 2001, ApJS, 136, 463
\\
Wakker, B.P., \& Savage, B.D. 2009, ApJS, 182, 378
\\
Werk, J.K., Prochaska, J.X., Thom, C., et al.\,2013, ApJS, 204, 17
\\
Werk, J.K., Prochaska, J.X., Tumlinson, J., et al.\,2014, ApJ, 792, 8
\\
White, S.D.M. \& Rees, M.J. 1978, MNRAS, 183, 341
\\
Wiersma, R.P.C., Schaye, J., \& Smith, B.D. 2009a, MNRAS, 393, 99
\\
Wiersma, R.P.C., Schaye, J., Theuns, T., Dalla Vecchia, C., \& Tornatore, L. 2009b, MNRAS, 399, 574
\\
Zwaan, M.A., van der Hulst, J.M., Briggs, F.H., Verheijen, M.A.W., \&
Ryan-Weber, E.V. 2005, MNRAS, 364, 1467
\\
Zhu,G. \& M'enard, B. 2013, ApJ, 770, 130
\\
\end{noindent}
\end{small}

%\end{thebibliography}

%%%%%%%%%%%%%%%%%%%%%%%%%%%%%%%%%%%%%%%%%%%%%%%%%%%%%%%%

\begin{appendix}

\section{Supplementrary Tables and Figures}

[Note: all velocity plots are available in the full
version at http://tucana.astro.physik.uni-potsdam.de/~prichter/si3cgm.pdf.]

%%%%%%%%%%%%%%%%%%%%%% TABLE A01 %%%%%%%%%%%%%%%%%%%%%%

\begin{table*}[h!]
\caption[]{COS QSO sample - part {\sc i}}
\begin{scriptsize}
\begin{tabular}{lllrrrr}
\hline
No. & QSO (COS name) & QSO (NED name) & $\alpha_{2000}$ & $\delta_{2000}$ & $z_{\rm QSO}$ & $\Delta z_{\rm abs}$ \\
\hline
        1 & 1ES1028+511              &   2MASX J10311847+5053358         &    157.83 &  50.89 &  0.360 &  0.095 \\
        2 & 1ES1553+113              &   [HB89] 1553+113                 &    238.93 &  11.19 &  0.360 &  0.098 \\
        3 & 1H-2129-624              &   2MASX J21362313-6224008         &    324.10 & -62.40 &  0.059 &  0.041 \\
        4 & 1SAXJ1032.3+5051         &   2MASSi J1032161+505120          &    158.07 &  50.86 &  0.173 &  0.098 \\
        5 & 2DFGRSS393Z082           &   GALEXASC J024500.80-300722.3    &     41.25 & -30.12 &  0.339 &  0.098 \\
        6 & 2DFGRSS394Z150$^{\rm a}$ &   GALEXASC J024613.27-302942.2    &     41.56 & -30.50 &  0.131 &  0.075 \\
        7 & 2E-1058.1+1102           &   [HB89] 1058+110                 &    165.20 &  10.77 &  0.423 &  0.098 \\
\hline
\end{tabular}
\\
$^{\rm a}$\,Spectrum shows no QSO characteristics; possibly wrong source (white dwarf) observed; not considered in study.
\end{scriptsize}
\end{table*}

%%%%%%%%%%%%%%%%%%%%%% TABLE A02 %%%%%%%%%%%%%%%%%%%%%%

\begin{table*}[h!]
\caption[]{COS QSO sample - part {\sc ii}}
\begin{scriptsize}
\begin{tabular}{lllrrrr}
\hline
No. & QSO (COS name) & QSO (NED name) & $\alpha_{2000}$ & $\delta_{2000}$ & $z_{\rm QSO}$ & $\Delta z_{\rm abs}$ \\
\hline
       77 & MCG+10-22-028            &   VV 487                          &    234.54 &  57.60 &  0.073 &  0.055 \\
       78 & MR2251-178               &   MR 2251-178                     &    343.52 & -17.58 &  0.064 &  0.046 \\
       79 & MRK-1044                 &   MRK 1044                        &     37.52 &  -9.00 &  0.016 &  0     \\
       80 & MRK-279                  &   UGC 08823                       &    208.26 &  69.31 &  0.030 &  0.012 \\
       81 & MRK106                   &   MRK 0106                        &    139.98 &  55.36 &  0.123 &  0.092 \\
       82 & MRK1513                  &   UGC 11763                       &    323.12 &  10.14 &  0.063 &  0.045 \\
       83 & MRK290                   &   MRK 0290                        &    233.97 &  57.90 &  0.030 &  0.010 \\
\hline
\end{tabular}
\end{scriptsize}
\end{table*}

%%%%%%%%%%%%%%%%%%%%%% TABLE A03 %%%%%%%%%%%%%%%%%%%%%%

\begin{table*}[h!]
\caption[]{COS QSO sample - part {\sc iii}}
\begin{scriptsize}
\begin{tabular}{lllrrrr}
\hline
No. & QSO (COS name) & QSO (NED name) & $\alpha_{2000}$ & $\delta_{2000}$ & $z_{\rm QSO}$ & $\Delta z_{\rm abs}$ \\
\hline
      153 & RBS144                   &   SARS 014.56153-51.50151         &     15.11 & -51.23 &  0.063 &  0.033 \\
      154 & RBS1892                  &   IRAS F22423-4707                &    341.33 & -46.87 &  0.200 &  0.088 \\
      155 & RBS563                   &   6dF J0438292-614759             &     69.62 & -61.80 &  0.069 &  0.045 \\
      156 & RXJ0439.6-5311           &   6dF J0439387-531131             &     69.91 & -53.19 &  0.243 &  0.098 \\
      157 & RXJ07145+7408            &   2MASSi J0714362+740810          &    108.65 &  74.14 &  0.371 &  0.095 \\
      158 & RXJ1230.8+0115           &   SDSS J123050.03+011522.6        &    187.71 &   1.26 &  0.117 &  0.098 \\
      159 & RXJ2154.1-4414           &   6dF J2154511-441406             &    328.71 & -44.23 &  0.344 &  0.098 \\
\hline
\end{tabular}
\end{scriptsize}
\end{table*}

%%%%%%%%%%%%%%%%%%%%%% TABLE A04 %%%%%%%%%%%%%%%%%%%%%%

\begin{table*}[h!]
\caption[]{COS QSO sample - part {\sc iv}}
\begin{scriptsize}
\begin{tabular}{lllrrrr}
\hline
No. & QSO (COS name) & QSO (NED name) & $\alpha_{2000}$ & $\delta_{2000}$ & $z_{\rm QSO}$ & $\Delta z_{\rm abs}$ \\
\hline
      229 & SDSSJ112244.89+575543.0  &   SDSS J112244.87+575543.1        &    170.69 &  57.93 &  0.907 &  0.098 \\
      230 & SDSSJ113327.78+032719.1  &   SDSS J113327.78+032719.1        &    173.37 &   3.46 &  0.525 &  0.098 \\
      231 & SDSSJ113457.62+255527.9  &   FBQS J113457.6+255527           &    173.74 &  25.92 &  0.710 &  0.098 \\
      232 & SDSSJ115758.72-002220.8  &   SDSS J115758.72-002220.7        &    179.49 &  -0.37 &  0.260 &  0.098 \\
      233 & SDSSJ120556.08+104253.8  &   [HB89] 1203+109                 &    181.48 &  10.72 &  1.089 &  0.093 \\
      234 & SDSSJ120720.99+262429.1  &   SDSS J120720.99+262429.1        &    181.84 &  26.41 &  0.322 &  0.098 \\
      235 & SDSSJ121037.56+315706.0  &   SDSS J121037.56+315706.0        &    182.66 &  31.95 &  0.389 &  0.098 \\
\hline
\end{tabular}
\end{scriptsize}
\end{table*}

%%%%%%%%%%%%%%%%%%%%%% TABLE A05 %%%%%%%%%%%%%%%%%%%%%%

\begin{table*}[h!]
\caption[]{Absorption-line systems\, - part {\sc i}}
\begin{scriptsize}
\begin{tabular}{llllrrrr}
\hline
No. & QSO (COS name) & $z_{\rm abs}$ & Ion & $W_{\rm r}^{\rm a}$ [m\AA] & $\Delta W_{\rm r}$ [m\AA] & log $N$ & $\Delta$(log $N$) \\
\hline
        1 & 2DFGRSS393Z082                &  0.004 & H\,{\sc i}   &     563  &         29 &        &        \\
        2 &                               &        & Si\,{\sc ii} &      33  &         12 &  12.49 &   0.25 \\
        3 &                               &        & Si\,{\sc iii}&      69  &         27 &  12.91 &   0.15 \\
        4 &                               &        & Si\,{\sc iv} &     <20  &            & <12.68 &        \\
        5 &                               &        & C\,{\sc ii}  &      53  &         18 &  13.53 &   0.23 \\
        6 &                               &        & C\,{\sc iv}  &          &            &        &        \\
        7 & 2MASS-J10512569+1247462       &  0.003 & H\,{\sc i}   &    1027  &        103 &        &        \\
\hline
\end{tabular}
\noindent
\\
$^{\rm a}$\,Used transitions: 
H\,{\sc i} $\lambda 1215$,
Si\,{\sc ii} $\lambda 1260$, 
Si\,{\sc iii} $\lambda 1206$, 
Si\,{\sc iv} $\lambda 1393$,
C\,{\sc ii} $\lambda 1334$, 
and C\,{\sc iv} $\lambda 1548$.
\end{scriptsize}
\end{table*}

%%%%%%%%%%%%%%%%%%%%%% TABLE A06 %%%%%%%%%%%%%%%%%%%%%%

\begin{table*}[h!]
\caption[]{Absorption-line systems\, - part {\sc ii}}
\begin{scriptsize}
\begin{tabular}{llllrrrr}
\hline
No. & QSO (COS name) & $z_{\rm abs}$ & Ion & $W_{\rm r}$ [m\AA] & $\Delta W_{\rm r}$ [m\AA] & log $N$ & $\Delta$(log $N$) \\
\hline
       73 & PG0832+251                    &  0.028 & H\,{\sc i}   &     325  &         18 &        &        \\
       74 &                               &        & Si\,{\sc ii} &      48  &         12 &  12.59 &   0.20 \\
       75 &                               &        & Si\,{\sc iii}&      40  &          8 &  12.35 &   0.16 \\
       76 &                               &        & Si\,{\sc iv} &     <13  &            & <12.25 &        \\
       77 &                               &        & C\,{\sc ii}  &      39  &          6 &  13.33 &   0.11 \\
       78 &                               &        & C\,{\sc iv}  &      34  &         12 &  13.03 &   0.20 \\
       79 & PG0832+251                    &  0.017 & H\,{\sc i}   &    1718  &         26 &        &        \\
\hline
\end{tabular}
\end{scriptsize}
\end{table*}

%%%%%%%%%%%%%%%%%%%%%% TABLE A07 %%%%%%%%%%%%%%%%%%%%%%

\begin{table*}[h!]
\caption[]{Absorption-line systems\, - part {\sc iii}}
\begin{scriptsize}
\begin{tabular}{llllrrrr}
\hline
No. & QSO (COS name) & $z_{\rm abs}$ & Ion & $W_{\rm r}$ [m\AA] & $\Delta W_{\rm r}$ [m\AA] & log $N$ & $\Delta$(log $N$) \\
\hline
      145 & PKS2005-489                   &  0.017 & H\,{\sc i}   &     311  &          8 &        &        \\
      146 &                               &        & Si\,{\sc ii} &     <12  &            & <11.98 &        \\
      147 &                               &        & Si\,{\sc iii}&      14  &          7 &  11.95 &   0.22 \\
      148 &                               &        & Si\,{\sc iv} &     <11  &            & <12.42 &        \\
      149 &                               &        & C\,{\sc ii}  &      10  &          5 &  12.80 &   0.22 \\
      150 &                               &        & C\,{\sc iv}  &      70  &         20 &  13.34 &   0.17 \\
      151 & PMNJ1103-2329                 &  0.084 & H\,{\sc i}   &     965  &         29 &        &        \\
\hline
\end{tabular}
\end{scriptsize}
\end{table*}

%%%%%%%%%%%%%%%%%%%%%% TABLE A08 %%%%%%%%%%%%%%%%%%%%%%

\begin{table*}[h!]
\caption[]{Absorption-line systems\, - part {\sc iv}}
\begin{scriptsize}
\begin{tabular}{llllrrrr}
\hline
No. & QSO (COS name) & $z_{\rm abs}$ & Ion & $W_{\rm r}$ [m\AA] & $\Delta W_{\rm r}$ [m\AA] & log $N$ & $\Delta$(log $N$) \\
\hline
      217 & SDSS-J150928.29+070235.7      &  0.079 & H\,{\sc i}   &     916  &         28 &        &        \\
      218 &                               &        & Si\,{\sc ii} &     140  &         35 &  13.07 &   0.11 \\
      219 &                               &        & Si\,{\sc iii}&     177  &         23 & >13.11 &        \\
      220 &                               &        & Si\,{\sc iv} &          &            &        &        \\
      221 &                               &        & C\,{\sc ii}  &      79  &         33 &  13.77 &   0.13 \\
      222 &                               &        & C\,{\sc iv}  &          &            &        &        \\
      223 & SDSS-J151237.05+012846.       &  0.029 & H\,{\sc i}   &    4319  &        351 &        &        \\
\hline
\end{tabular}
\end{scriptsize}
\end{table*}

%%%%%%%%%%%%%%%%%%%%%% TABLE A09 %%%%%%%%%%%%%%%%%%%%%%

\begin{table*}[h!]
\caption[]{Absorption-line systems\, - part {\sc v}}
\begin{scriptsize}
\begin{tabular}{llllrrrr}
\hline
No. & QSO (COS name) & $z_{\rm abs}$ & Ion & $W_{\rm r}$ [m\AA] & $\Delta W_{\rm r}$ [m\AA] & log $N$ & $\Delta$(log $N$) \\
\hline
      289 & SDSSJ110406.94+314111.4       &  0.062 & H\,{\sc i}   &     538  &         28 &        &        \\
      290 &                               &        & Si\,{\sc ii} &     <18  &            & <12.16 &        \\
      291 &                               &        & Si\,{\sc iii}&      25  &          8 &  12.17 &   0.19 \\
      292 &                               &        & Si\,{\sc iv} &     <21  &            & <12.46 &        \\
      293 &                               &        & C\,{\sc ii}  &      19  &          8 &  13.00 &   0.20 \\
      294 &                               &        & C\,{\sc iv}  &     279  &         34 & >14.04 &        \\
      295 & SDSSJ111239.11+353928.2       &  0.025 & H\,{\sc i}   &     955  &         52 &        &        \\
\hline
\end{tabular}
\end{scriptsize}
\end{table*}

%%%%%%%%%%%%%%%%%%%%%% TABLE A10 %%%%%%%%%%%%%%%%%%%%%%

\begin{table*}[h!]
\caption[]{Absorption-line systems\, - part {\sc vi}}
\begin{scriptsize}
\begin{tabular}{llllrrrr}
\hline
No. & QSO (COS name) & $z_{\rm abs}$ & Ion & $W_{\rm r}$ [m\AA] & $\Delta W_{\rm r}$ [m\AA] & log $N$ & $\Delta$(log $N$) \\
\hline
      361 & SDSSJ140732.25+550725.6       &  0.005 & H\,{\sc i}   &    3422  &         89 &        &        \\
      362 &                               &        & Si\,{\sc ii} &     550  &         17 & >13.93 &        \\
      363 &                               &        & Si\,{\sc iii}&     769  &         21 & >13.85 &        \\
      364 &                               &        & Si\,{\sc iv} &     294  &         27 & >13.67 &        \\
      365 &                               &        & C\,{\sc ii}  &     610  &         19 & >14.75 &        \\
      366 &                               &        & C\,{\sc iv}  &          &            &        &        \\
      367 & SDSSJ141542.90+163413.8       &  0.008 & H\,{\sc i}   &    3267  &        376 &        &        \\
\hline
\end{tabular}
\end{scriptsize}
\end{table*}

%%%%%%%%%%%%%%%%%%%%%% TABLE A11 %%%%%%%%%%%%%%%%%%%%%%

\begin{table}[h!]
\caption[]{Si\,{\sc iii} candidate systems$^{\rm a}$}
\begin{tabular}{llc}
\hline
No. & QSO (COS name) & $z_{\rm abs}$ \\
\hline
        1 & 2MASS-J10131797+0500342  &       0.046 \\
        2 & 4C--01.61                &       0.038 \\
        3 & IRAS01003-2238           &       0.040 \\
        4 & MRK1513                  &       0.028 \\
        5 & PG1112+431               &       0.018 \\
        6 & PG1202+281               &       0.051 \\
        7 & PHL1811                  &       0.012 \\
        8 & RXJ0439.6-5311           &       0.057 \\
        9 & SDSSJ014143.20+134032.0  &       0.003 \\
       10 & SDSSJ024250.85-075914.2  &       0.005 \\
       11 & SDSSJ040148.98-054056.5  &       0.029 \\
       12 & SDSSJ094331.61+053131.4  &       0.086 \\
       13 & SDSSJ101622.60+470643.3  &       0.057 \\
       14 & SDSSJ102218.99+013218.8  &       0.067 \\
       15 & SDSSJ122018.43+064119.6  &       0.086 \\
       16 & SDSSJ150455.56+564920.3  &       0.036 \\
       17 & SDSSJ155048.29+400144.9  &       0.077 \\
       18 & SDSSJ155304.92+354828.6  &       0.028 \\
       19 & SDSSJ161649.42+415416.3  &       0.089 \\
       20 & VV2006-J015953.0+134554  &       0.004 \\

\hline
\end{tabular}
\noindent
\\
{\small $^{\rm a}$\,detected only in Si\,{\sc iii} and H\,{\sc i} Ly\,$\alpha$, but
not confirmed in other metal lines.}
\\
\end{table}

%%%%%%%%%%%%%%%%%%%%%% TABLE A12 %%%%%%%%%%%%%%%%%%%%%%

\begin{table*}[h!]
\caption[]{Comparison with (revised) Danforth et al.\,(2016) absorber list}
\begin{tiny}
\begin{tabular}{lrl}
\hline
QSO (COS name) & $z_{\rm abs}$ & Comment \\
\hline
\multicolumn{3}{c}{Danforth detections without Richter counterpart} \\
\hline
3C57           &  0.077417 &  many line blends \\
3C57           &  0.089742 &  candidate metal lines are offset from Ly\,$\alpha$ \\
MR2251-178     &  0.010725 &  candidate metal lines are offset from Ly\,$\alpha$ \\
RXJ1230.8+0115 &  0.071728 &  no evidence for metal absorption \\
\hline
\multicolumn{3}{c}{Richter candidate system without Danforth counterpart} \\
\hline
RXJ0439.6-5311 &  0.057400 &  possible Si\,{\sc iii} absorber present, but blend with intrinsic Ly\,$\beta$ \\
\hline
\end{tabular}
\end{tiny}
\end{table*}

\end{appendix}
\end{document}